\documentclass[final,3p,times]{elsarticle}

\usepackage{amssymb}
\usepackage{mathtools}
\usepackage{hyperref}
\usepackage{braket}
\usepackage{epsfig}
\usepackage{mwe}
\usepackage{subfig}
\usepackage{float}

\journal{Physica A}

\begin{document}

\begin{frontmatter}

\title{Long Range Quantum Coherence, Quantum \& Classical Correlations \\in Heisenberg XX Chain}

\author[label1]{Zakaria Mzaouali\corref{cor1}}
\address[label1]{ESMaR, Faculty of Sciences, Mohammed V University, Av. Ibn Battouta, B.P. 1014, Agdal, Rabat, Morocco}

\cortext[cor1]{Corresponding author}
\ead{zakaria.mzaouali@um5s.net.ma}

\author[label1]{Morad El Baz}

\begin{abstract}
A comparative study of pairwise quantum coherence, quantum and classical correlations is addressed for non-nearest spin pairs of the 1D Heisenberg spin-$\frac{1}{2}$ XX chain. Following the Jordan-Wigner mapping, we diagonalise the hamiltonian of the chain and we check this procedure numerically as well. Using the ``Pauli basis expansion" formalism we get the pairwise quantities studied in this work at any distance. We then, show the role of quantum correlations in revealing quantum phase transitions, the robustness of quantum discord to the temperature and the dominance of quantum correlations over their classical counterpart in the magnetic and thermal interval in quantum spin chains. We conclude the paper by shedding light from a resource-driven point of view on the new born quantity ``quantum coherence" where we discuss its role in detecting quantum phase transitions being a long-range quantity, and how it outclasses the usual quantum correlations measures in the robustness against the temperature, which indicates potential uses in the framework of quantum information processing.
\end{abstract}

\begin{keyword}
Entanglement, Quantum Discord, Classical Correlations, Quantum Coherence, Spin Chains, Quantum Phase Transitions
\end{keyword}

\end{frontmatter}


\section{Introduction}
Correlations, the key elements to understand many-body systems can be classified into classical and quantum types. The existence of quantum correlations were first pointed out in 1935 by E. Schr\"odinger \cite{schrodinger1935} in the context of non-separable (i.e entangled) states. Since then, much effort was devoted in studying the nature of entanglement \cite{entanglement_review,epr} especially in the framework of resource theory in quantum information processing, through quantum teleportation \cite{teleportation}, quantum communication \cite{qcomunication} and quantum computation \cite{qcomputation}.

The quantum information approach to condensed matter physics has been very fruitful at giving new a perspective to understand the collective phenomena in many-body systems. Indeed, since the early 2000’s various measures of quantum entanglement  have been employed to characterize the features of the ground and excited states of quantum matter \cite{rosario,ent_nielsen}. In this sense, quantum spin systems play an essential role in these developments as they describe the effective interactions in a collection of physical systems \cite{quantum_magnetism} like quantum Hall systems, high-temperature superconductors, heavy fermions and magnetic compounds. Furthermore, systems that can be described by interacting spins are interesting because they manifest quantum fluctuations and can be realised by a variety of physical approaches \cite{creation_qsc_1,creation_qsc_2}.

The study of spin entanglement in quantum spin chains began in 2000’s. Since then, many types of entanglement at both zero and finite temperatures have been widely studied in various spin models \cite{rosario}. However, entanglement measures fails at capturing other forms of non-classical correlations \cite{discord_vedral,quantum_correlation} that emerge “for free” in the ground and thermal states of condensed matter models and that can be exploited as resources for quantum technologies. As a matter of fact, quantum discord \cite{discord} is the most effective measure of quantum correlations beyond entanglement, and it has been heavily studied in spin chains at both zero \cite{discord_zero1,discord_zero2,discord_zero3} and finite temperature \cite{discord_zero3,discord_finite1,discord_finite2}. As entanglement plays an important role in identifying quantum phase transitions \cite{ent_nielsen,rosario_qpt}, quantum discord has received much attention in this regard as well \cite{discord_qpt1,discord_qpt2}. Moreover, it was applied in several contexts like open quantum systems \cite{discord_oqs}, quantum dynamics \cite{discord_dynamics} and even biophysics \cite{discord_bio}.

Recently, the concept of quantum coherence has received much attention in the quantum information community as it plays an essential role in phenomena like quantum interference, bipartite and multipartite entanglement \cite{coherence_as_resource}. Various schemes were proposed for detecting coherence \cite{old_coherence1,old_coherence2}, but it was never quantified in the language of quantum information theory until the seminal work of Baumgratz, Cramer and Plenio \cite{baumgratz} in which they constructed a quantitative theory that captures the resource character of coherence in a mathematically rigorous fashion. Such developments led to number of applications using coherence as a basic ingredient in various fields such as quantum communication \cite{coherence_communication} and in farther other arenas, such as thermodynamics \cite{coherence_thermo} and even certain branches of biology \cite{coherence_bio}.\\
Few works were dedicated to study quantum coherence in condensed matter systems \cite{coherence_spin1,coherence_spin2,coherence_spin3,coherence_spin4}. In fact, the investigations that were carried out in spin chains like the XY model had the sole purpose of revealing the connection between quantum coherence and quantum phase transitions. These studies has shown the role played by coherence in detecting important features like critical points, but it is still early to say how efficient quantum coherence is in detecting quantum phase transitions as the field needs more models and measures to investigate these connections.

Motivated by these developments, the aim of this paper is to study and compare the behavior of  non-nearest quantum coherence, quantum and classical correlations in an infinite 1D spin-$\frac{1}{2}$ XX chain in the presence of a magnetic field. This is an analytically solvable model by means of the Jordan-Wigner transformation which we check numerically as well. Also this model has well-known physical properties hence it is a suitable ground for studying the interface between quantum information theory and condensed matter physics.

This article is organised as follows. In the next section we introduced the model, the analytical and numerical diagonalisation process by the Jordan-Wigner mapping. In section \ref{section3}, we introduce the two-sites density matrix and the various measures of correlations considered in this paper. In section \ref{section4}, we discuss our results on non-nearest quantum coherence, quantum and classical correlations in the XX chain. A conclusion and the summary of the results are presented in \ref{section5}.
\section{The Model}
The Hamiltonian of the Heisenberg XX chain describing a set of localized spin-$\frac{1}{2}$ particles interacting with nearest-neighbors exchanging coupling on a 1D lattice in an external magnetic field is given by :
    \begin{equation}
        \mathcal{H_{XX}}=J\sum_{i=1}^N \big(S_i^xS_{i+1}^x+S_i^yS_{i+1}^y\big)-h\sum_{i=1}^NS_i^z,
        \label{eq1}
    \end{equation}
where $S_i$ is the spin-$\frac{1}{2}$ operator on site $i$. A positive (negative) exchange coupling $J$ favors anti-ferromagnetic (ferromagnetic) ordering of the spins, and $h$ is the external magnetic field which interacts only with the $z$-component of the spins.\\

The pecularity of the XX model is the possibility of exact diagonalisation by using a method that is very different from the Bethe Ansatz \cite{bethe}, and can be verified numerically as well. This method consists in mapping each spin operator to a fermion operator, following the Jordan-Wigner transformation. For simplicity, we first introduce the spin ladder operators :
    \begin{equation}
        S^{\pm}=S^x\pm i S^y 
        \quad\text{ with}\quad
        S^+=\begin{pmatrix}
        0&1\\0&0
        \end{pmatrix}
        \quad\text{,}\quad
        S^-=\begin{pmatrix}
        0&0\\1&0
        \end{pmatrix}.
        \label{eq2}
    \end{equation}
    
In 1928, Jordan and Wigner showed \cite{jordan_wigner} that the spin ladder operators can be represented exactly by the fermion operators ($c_i$'s) with the following mapping :
    \begin{equation}
        S_i^+=c_i^+e^{\displaystyle{-j\pi\sum_{l=1}^{i-1}c_l^+c_l}}
        \quad\text{,}\quad
        S_i^-=e^{\displaystyle{j\pi\sum_{l=1}^{i-1}c_l^+c_l}}c_i
        \quad\text{,}\quad
        S_i^z=c_i^+c_i-\frac{1}{2},
        \label{eq3}
    \end{equation}
where $j$ is the imaginary unit.\\
Rewriting the Hamiltonian (\ref{eq1}) in terms of the ladder operators (\ref{eq2}) and by performing the  transformation (\ref{eq3}), the Hamiltonian is readily shown to take
the form :
    \begin{equation}
        \mathcal{H_{XX}}=J\sum_{i=1}^N\left(c_i^+c_{i+1}+c_{i+1}^+c_i\right)-h\sum_{i=1}^N\left(c_i^+c_i-\frac{1}{2} \right).
        \label{eq4}
    \end{equation}
    
Since The XX model is translation invariant, we can introduce the Fourier transform of the fermion operators : 

    \begin{equation}
        d_k=\frac{1}{\sqrt{N}} \sum_{i=1}^N e^{\displaystyle{-jik}}c_i
        \quad\text{and}\quad
        d_k^+=\frac{1}{\sqrt{N}}\sum_{i=1}^N e^{\displaystyle{jik}}c_i^+.
    \end{equation}
    
The final diagonalised form of the Hamiltonian of the XX-chain written in terms of the
spinless fermion creation and annihilation operators is : 

    \begin{equation}
    \mathcal{H_{XX}}=\sum_k \epsilon(k)d_k^+d_k
    \quad\text{with}\quad
    \epsilon(k)=J\cos(k)-h.
    \end{equation}

In the rest of this paper the coupling constant $J$ will be equal to unity. The Jordan-Wigner transformation can be verified using ``QuSpin'' \cite{quspin1,quspin2} a Python package for dynamics and exact diagonalisation of quantum many-body systems.
    \begin{figure}[h]
        \centering
        \includegraphics[scale=0.5]{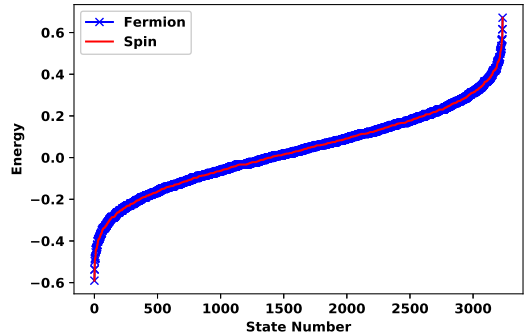}
        \caption{The Spectrum of the XX-model in the spin and fermion representation}
        \label{fig:1}
    \end{figure}
\\
In Figure.\ref{fig:1} we plot for a chain of 15 spins, $J=1$ and $h=1$ with periodic boundary conditions the energy spectrum of the XX chain with respect to the number of states which is reduced due to the symmetries of the model. The red\footnotemark solid line corresponds to the eigenenergies of the usual spin representation of the XX model (\ref{eq1}), while the blue\footnotemark[\value{footnote}] crosses are the eigenvalues of the fermionic Hamiltonian (\ref{eq4}) obtained by the Jordan-Wigner mapping. We see that for each spin state there is a corresponding fermion state, which means that the two representations match exactly, as predicted.
\footnotetext{color on the online version}
\section{Correlation Measures in Spin Chains}
\label{section3}
Quantifying correlations that emerge in a given quantum system requires the knowledge of the density matrix of the state characterizing the system. Here we are interested in studying the bipartite correlations between two spins $i$ and $j$ distant by some lattice spacing $m$, where $i=j+m$. Using Pauli basis expansion \cite{wells}, the two sites density matrix $\rho_{i,i+m}$ is given by :
    \begin{equation}
        \rho_{i,i+m}=\frac{1}{4}\sum_{\alpha,\beta=0}^3 p_{\alpha \beta}  \sigma_i^{\alpha}\otimes\sigma_{i+m}^{\beta},
    \end{equation}
with $p_{\alpha \beta}=\langle\sigma_i^{\alpha}\sigma_{i+m}^{\beta}\rangle$, and $\sigma_i^\alpha$ are the Pauli matrices, $\alpha=0,1,2,3$. In terms of Pauli ladder operators $\sigma^{\pm}=\frac{1}{2}(\sigma^x\pm i\sigma^y)$, we can write in the computational basis $\{\ket{\uparrow\uparrow},\ket{\uparrow\downarrow},\ket{\downarrow\uparrow},\ket{\downarrow\downarrow}\}$, the density matrix of two spins as follows : 
    \begin{equation}
        \rho_{i,i+m}=
        \begin{pmatrix}
        \langle P_i^{\uparrow}P_{i+m}^{\uparrow} \rangle&\langle P_i^{\uparrow}\sigma_{i+m}^-\rangle&\langle \sigma_i^-P_{i+m}^{\uparrow}\rangle&\langle \sigma_i^-\sigma_{i+m}^-\rangle\\
        \langle P_i^{\uparrow}\sigma_{i+m}^+\rangle&\langle P_i^{\uparrow}P_{i+m}^{\downarrow} \rangle&\langle \sigma_i^-\sigma_{i+m}^+\rangle&\langle \sigma_i^-P_{i+m}^{\downarrow}\rangle\\
        \langle \sigma_i^+P_{i+m}^{\uparrow}\rangle&\langle \sigma_i^+\sigma_{i+m}^-\rangle&\langle P_i^{\downarrow}P_{i+m}^{\uparrow} \rangle&\langle P_i^{\uparrow}\sigma_{i+m}^{-}\rangle\\
        \langle \sigma_i^+\sigma_{i+m}^+\rangle&\langle \sigma_i^+P_{i+m}^{\downarrow}\rangle&\langle P_i^{\uparrow}\sigma_{i+m}^{+}\rangle&\langle P_i^{\downarrow}P_{i+m}^{\downarrow} \rangle
        \end{pmatrix},
        \label{eq5}
    \end{equation}

where $P^{\uparrow}=\frac{1}{2}(1+\sigma^z)$ and $P^{\downarrow}=\frac{1}{2}(1-\sigma^z)$. The brackets denote the ground-state and thermodynamic average values at zero and finite temperatures. \\

The Hamiltonian (\ref{eq1}) has several symmetries that reduces the number of non-zero elements of the density matrix (\ref{eq5}). For instance, translation invariance means that $\rho_{i,i+m}=\rho_{i+m,i}$, therefore we have $p_{\alpha\beta}=p_{\beta\alpha}$. The global phase-flip symmetry implies that the commutator $\left[\sigma_i^z\sigma_{i+m}^z,\rho_{i,i+m}\right]=0$, then the coefficients $p_{01}=p_{10}$, $p_{02}=p_{20}$, $p_{13}=p_{31}$ and $p_{23}=p_{32}$ must be zero. Furthermore, the $z$-component of the total magnetization commutes with the Hamiltonian (\ref{eq1}) $\left[ \sum_{i} \sigma_i^z,\mathcal{H_{XX}} \right]=0$ which implies that a coherent superposition of basis states $\ket{\uparrow\uparrow}$ and $\ket{\downarrow\downarrow}$ can not be possible for any two spins. Then the coefficient $p_{30}$ and $p_{03}$ corresponding to the matrix elements $\ket{\uparrow\uparrow}\bra{\downarrow\downarrow}$ and $\ket{\downarrow\downarrow}\bra{\uparrow\uparrow}$ are zero. The density matrix (\ref{eq5}) reduces to : 

    \begin{equation}
    \rho_{i,i+m}=
        \begin{pmatrix}
        X_{i,i+m}^+&0&0&0\\
        0&Y_{i,i+m}^+&Z_{i,i+m}^*&0\\
        0&Z_{i,i+m}&Y_{i,i+m}^-&0\\
        0&0&0&X_{i,i+m}^-
        \end{pmatrix}.
        \label{eq6}
    \end{equation}

\noindent In terms of the occupation number operator $n_i=c_i^+c_i$, the elements of the density matrix are given by :
    \begin{equation}
        \begin{aligned}
        X_{i,i+m}^+&=\langle n_in_{i+m} \rangle,
        \quad\text{}\quad
        X_{i,i+m}^-=1- \langle n_i \rangle - \langle n_{i+m} \rangle+\langle n_in_{i+m} \rangle, \\
        \quad\text{}\quad
        Y_{i,i+m}^+&=\langle n_i \rangle-\langle n_i n_{i+m}\rangle,
        \quad\text{}\quad
        Y_{i,i+m}^-=\langle n_{i+m} \rangle-\langle n_{i+m}n_i\rangle, \\
        Z_{i,i+m}&=\langle c_i^+\Big(\prod_{k=i}^{i+m-1}\{1-2c_k^+c_k\}\Big)c_{i+m} \rangle.
        \end{aligned}
        \label{eq7}
    \end{equation}

In the coming sections we investigate various contributions of correlations at long distances of $m$ in the XX chain. We study the correlations between second nearest $(2N)$ $m=2$, third nearest $(3N)$ $m=3$ and the fourth neighbor $(4N)$ $m=4$. Then, by using Wick's theorem \cite{wick} the set of equations (\ref{eq7}) reduces to :
\begin{equation}
    \begin{aligned}
        X_{i,i+m}^+&=f_0^2-f_m^2,
        \quad\text{}\quad
        X_{i,i+m}^-=1-2f_0+f_0^2-f_m^2,\\
        Y_{i,i+m}^+&=Y_{i,i+m}^-=f_0-f_0^2+f_m^2,
    \label{eq_diag}
    \end{aligned}
\end{equation}
\noindent while for the element $Z_{i,i+m}$ we have
\begin{equation}
    \begin{aligned}
    \quad\text{for $m=2$,}\quad
    Z_{i,i+2}&=f_2-2f_0f_2+2f_1^2,\\
    \quad\text{for $m=3$,}\quad
    Z_{i,i+3}&=4(f_1^3-2f_0f_1f_2+f_2^2f_1+f_0^2f_3-f_1^2f_3+f_1f_2-f_0f_3)+f_3, \\
    \quad\text{for $m=4$,}\quad
    Z_{i,i+4}&=8(f_1^4-3f_0f_1^2f_2+2f_1^2f_2^2+2f_0^2f_1f_3+f_0^2f_2^2-f_2^4-2f_0f_1f_2f_3\\&+2f_1f_2^2f_3-2f_1^3f_3+f_1^2f_3^2-f_0f_2f_3^2-f_0^3f_4+2f_0f_1^2f_4-2f_1^2f_2f_4\\&+f_0f_2^2f_4)+4(3f_1^2f_2-2f_0f_2^2-4f_0f_1f_3+2f_1f_2f_3+3f_0^2f_4-2f_1^2f_4\\&+f_2f_3^2-f_2^2f_4)+2(2f_1f_3-3f_0f_4+f_2^2)+f_4.
    \label{eqz}
    \end{aligned}
\end{equation}
\noindent For the non-negative integer number $m$:
    \begin{equation}
        f_m=\frac{1}{\pi}\int_{0}^{\pi}\cos(km)g(k)dk,
        \label{eq11}
    \end{equation}
\noindent where $g(k)=\frac{1}{1+\displaystyle{e^{\beta \epsilon(k)}}}$ is the Fermi-Dirac distribution, $\beta=\frac{1}{k_B T}$ and the Boltzmann constant is taken equal to unity.
\subsection{Entanglement}
One candidate to study the bipartite properties of non-local quantum correlations of a quantum state is by means of the concurrence \cite{concurrence1,concurrence2} defined as follows :
    \begin{equation}
    C(\rho)=\max\{0,\lambda_1-\lambda_2-\lambda_3-\lambda_4\},
    \label{eq12}
    \end{equation}
where the $\lambda_i$'s are the square roots of the eigenvalues of the matrix $\rho\Tilde{\rho}$ in decreasing order, with $\Tilde{\rho}$ being a transformed matrix of $\rho$ i.e., $\Tilde{\rho}=(\sigma_y \otimes \sigma_y) \rho^* (\sigma_y \otimes \sigma_y)$. In the case of the two sites density matrix $\rho_{i,i+m}$ (\ref{eq6}) considered in this paper, the concurrence then can be calculated from the local-density's, hopping term, and site-site correlations of the fermions. Thus, Eq.(\ref{eq12}) takes the following form :
    \begin{equation}
        C(\rho_{i,i+m})=\max\Big\{0,2\Big(|Z_{i,i+m}|-\sqrt{X_{i,i+m}^+X_{i,i+m}^-}\Big)\Big\},
    \end{equation}
where $Z_{i,i+m},X_{i,i+m}^+$ and $X_{i,i+m}^-$ are given by the set of equations (\ref{eq_diag}) and (\ref{eqz}).
\subsection{Total Quantum \& Classical Correlations}
A bipartite quantum state can carry quantum and classical correlations \cite{discord_vedral}, entanglement quantifiers like the concurrence fails at detecting quantum correlations other than the non-local ones. To measure the total quantum correlations in a given bipartite quantum system $\rho_{AB}$ we use Quantum Discord (QD) a measure introduced by Olivier \& Zurek \cite{discord}, and independently by Henderson \& Vedral \cite{discord_vedral}. It is based on the difference between the total correlations and classical correlations (CC):
    \begin{equation}
        \mathcal{QD(\rho_{AB})}=\mathcal{I(\rho_{A:B})}-\mathcal{CC(\rho_{AB})},
    \end{equation}
  
\noindent where the total correlations are quantified by the quantum mutual information :
    \begin{equation}
        \mathcal{I(\rho_{A:B})}=\mathcal{S(\rho_{A})}+\mathcal{S(\rho_{B})}-\mathcal{S(\rho_{AB})},
        \label{eq8}
    \end{equation}
    
\noindent with $S(\rho)=-\text{tr}(\rho\log_2{\rho})$ being the von-Neumann entropy.\\

Classical correlations are defined as the maximization of $\mathcal{I}(\rho_{A:B}|\{ \prod_k^B\})$ over the set of positive operator valued measurements (POVM) $\{\prod_k^B\}$ on subsystem B, and is given by :

    \begin{equation}
        \mathcal{CC(\rho_{AB})}=\sup_{\{\prod_k^B\}} \mathcal{I}(\rho_{A:B}|\{\prod_k^B\}).
    \end{equation}
It reduces to :
    \begin{equation}
        \mathcal{CC(\rho_{AB})}=S(\rho_{A})-\min_{\prod_k^B}S(\rho_{B}|\prod_k^B),
        \label{eq9}
    \end{equation}
where $S(\rho_{AB}|\prod_k^B)=\sum_k p_k S(\rho_k^B)$ is the conditional entropy based on the measurement $\prod_k^B$. $p_k=\text{Tr}\big[(\prod_k^B \otimes I_B)\rho_{AB}(\prod_k^B \otimes I_B)\big]$, and $\rho_{B}^k=\displaystyle{\frac{(\prod_k^B \otimes I_B)\rho_{AB}(\prod_k^B \otimes I_B)}{p_k}}$ are the probability and the state for a measurement outcome $k$.\\
From Eq.(\ref{eq8}) and Eq.(\ref{eq9}) quantum discord takes the form :
    \begin{equation}
        \mathcal{QD(\rho_{AB})}=\mathcal{S(\rho_{A})}-\mathcal{S(\rho_{AB})}+\min_{\prod_k^B}S(\rho_{B}|\prod_k^B).
    \end{equation}

The difficulty in evaluating QD rises from the minimization procedure of the conditional entropy, and obtaining an analytical expression of $CC(\rho_{AB})$ is not an easy task for general states. However, following the method of C.Z Wang and al in \cite{xmatrix} an explicit expression for $CC(\rho_{AB})$ and $QD(\rho_{AB})$ is available for X-states described by density matrices of the form :
    \begin{equation}
    \rho_{AB}=
        \begin{pmatrix}
         \rho_{11}&0&0&\rho_{14}\\
         0&\rho_{22}&\rho_{23}&0\\
         0&\rho_{32}&\rho_{33}&0\\
         \rho_{41}&0&0&\rho_{44}
        \end{pmatrix}.
    \end{equation}.
    
\noindent We obtain $CC(\rho_{AB})$ and $QD(\rho_{AB})$ as follows :
\begin{equation}
            \begin{aligned}
            CC(\rho_{AB})=\max\{CC_1,CC_2\},\\
            QD(\rho_{AB})=\min\{QD_1,QD_2\},
        \end{aligned}
\end{equation}
where 
\begin{equation}
            \begin{aligned}
            CC_j&=H(\rho_{11}+\rho_{22})-D_j,\\
            QD_j&=H(\rho_{11}+\rho_{33})+\sum_{k=1}^4 \lambda_k \log_2(\lambda_k)+D_j,
        \end{aligned}
\end{equation}

and
\begin{equation}
            \begin{aligned}
            D_1&=H(w), D_2=-\sum_{j=1}^4 \rho_{jj} \log_2(\rho_{jj})-H(\rho_{11}+\rho_{33}),\\
            w&=\displaystyle{\frac{1+\sqrt{[1-2(\rho_{33}+\rho_{44})]^2+4(|\rho_{14}|+|\rho_{23}|)^2}}{2}},
        \end{aligned}
\end{equation}
$H(x):=-x\log_2(x)-(1-x)\log_2(1-x)$ being the binary Shannon entropy and $\lambda_k$ are the eigenvalues of the matrix $\rho_{AB}$.\\

Then, for the two sites density matrix (\ref{eq6}) considered in this paper, the analytical expression for $CC(\rho_{i,i+m})$ and $QD(\rho_{i,i+m})$ are given by :
\begin{equation}
            \begin{aligned}
            CC_1(i,i+m)&=H(X_{i,i+m}^++Y_{i,i+m}^+)-H\Bigg(\frac{1+\sqrt{[1-2(Y_{i,i+m}^-+X_{i,i+m}^-)]^2+4|Z_{i,i+m}|^2}}{2}\Bigg),\\
            CC_2(i,i+m)&=H(X_{i,i+m}^++Y_{i,i+m}^+)+\{X_{i,i+m}^+\log_2(X_{i,i+m}^+)+Y_{i,i+m}^+\log_2(Y_{i,i+m}^+)\\&+Y_{i,i+m}^-\log_2(Y_{i,i+m}^-)+X_{i,i+m}^-\log_2(X_{i,i+m}^-)\}+H(X_{i,i+m}^++Y_{i,i+m}^-),\\
            QD_1(i,i+m)&=H(X_{i,i+m}^++Y_{i,i+m}^-)
            +\{X_{i,i+m}^+\log_2{X_{i,i+m}^+}+X_{i,i+m}^-\log_2{X_{i,i+m}^-}\\&+(Y_{i,i+m}^+-|Z_{i,i+m}|)\log_2{(Y_{i,i+m}^+-|Z_{i,i+m}|)}+(Y_{i,i+m}^-+|Z_{i,i+m}|)\log_2{(Y_{i,i+m}^-+|Z_{i,i+m}|)}\}\\&
            +H\Bigg(\frac{1+\sqrt{[1-2(Y_{i,i+m}^-+X_{i,i+m}^-)]^2+4|Z_{i,i+m}|^2}}{2}\Bigg),\\
            QD_2(i,i+m)&=H(X_{i,i+m}^++Y_{i,i+m}^-)
            +\{X_{i,i+m}^+\log_2{X_{i,i+m}^+}+X_{i,i+m}^-\log_2{X_{i,i+m}^-}\\&+(Y_{i,i+m}^+-|Z_{i,i+m}|)\log_2{(Y_{i,i+m}^+-|Z_{i,i+m}|)}+(Y_{i,i+m}^-+|Z_{i,i+m}|)\log_2{(Y_{i,i+m}^-+|Z_{i,i+m}|)}\}\\&
            -\{X_{i,i+m}^+\log_2(X_{i,i+m}^+)+Y_{i,i+m}^+\log_2(Y_{i,i+m}^+)+Y_{i,i+m}^-\log_2(Y_{i,i+m}^-)+X_{i,i+m}^-\log_2(X_{i,i+m}^-)\}\\&
            -H(X_{i,i+m}^++Y_{i,i+m}^-),
        \end{aligned}
\end{equation}
where $X_{i,i+m}^+$,$X_{i,i+m}^-$,$Y_{i,i+m}^+$,$Y_{i,i+m}^-$ and $Z_{i,i+m}$ are given by the set of equations (\ref{eq_diag}) and (\ref{eqz}). 
\section{Quantum Coherence}
Arising from quantum superposition, quantum coherence is a fundamental feature of quantum mechanics and plays a central role in the field of quantum information science. In the framework of resource theory, Baumgratz \textit{et al.} \cite{baumgratz} proposed a formalism for measuring quantum coherence which triggered the interest of many researchers and a variety of coherence measures emerged since then. In general, all these coherence measures can be classified into either the entropic or the geometric class of measures \cite{coherence_distri}. The classification depends on whether the measure is based on the entropy functional or has a metric nature which implies a geometric structure. Interestingly, there is a measure which combines both entropic and metric properties introduced in \cite{coherence_distri}, and is based on the quantum version of the Jensen-Shannon divergence \cite{jsd} $\mathcal{J}(\rho,\rho_d)$ defined as :
    \begin{equation}
        \mathcal{J}(\rho,\rho_d) = \mathcal{S}\Big(\frac{\rho+\rho_d}{2}\Big)-\frac{\mathcal{S}(\rho)}{2}-\frac{\mathcal{S}(\rho_d)}{2}.
    \end{equation}
The  quantum coherence is then simply the square root of the quantum Jensen-Shannon divergence : 
\begin{equation}
        \begin{aligned}
            \mathcal{QC}&=\sqrt{\mathcal{J}(\rho,\rho_d)}\\
            &=\sqrt{\mathcal{S}\Big(\frac{\rho+\rho_d}{2}\Big)-\frac{\mathcal{S}(\rho)}{2}-\frac{\mathcal{S}(\rho_d)}{2}},
    \end{aligned}
\end{equation}
Where $\rho$ is the density matrix of the state and $\rho_d$ that of the closest incoherent state under the distance measure and $S(.)$ is the von Neumann entropy. The closest incoherent state is taken to be the denstiy matrix where all the off-diagonal elements of $\rho$ are zero \cite{coherence_distri}.\\

Then, in terms of $Y_{i,i+m}^+$,$Y_{i,i+m}^-$ and $Z_{i,i+m}$ given by the set of equations (\ref{eq_diag}) and (\ref{eqz}), the analytical expression of quantum coherence for the XX chain described by the density matrix (\ref{eq6}) is : 
\begin{equation}
    \begin{aligned}
            \mathcal{QC}(\rho_{i,i+m})&=\Bigg[\Big\{
            -\Big(Y_{i,i+m}^+-\frac{|Z_{i,i+m}|}{2}\Big)\log_2{\Big(Y_{i,i+m}^+-\frac{|Z_{i,i+m}|}{2}\Big)}-\Big( Y_{i,i+m}^-+\frac{|Z_{i,i+m}|}{2}\Big)\\&\log_2{\Big(Y_{i,i+m}^-+\frac{|Z_{i,i+m}|}{2}\Big)}\Big\}-\frac{1}{2}\Big\{-(Y_{i,i+m}^+-|Z_{i,i+m}|)\log_2{(Y_{i,i+m}^+-|Z_{i,i+m}|)}\\&-(Y_{i,i+m}^-+|Z_{i,i+m}|)\log_2{(Y_{i,i+m}^-+|Z_{i,i+m}|)} \Big\}-\frac{1}{2}\Big\{-Y_{i,i+m}^+\log_2{Y_{i,i+m}^+}\\& -Y_{i,i+m}^-\log_2{Y_{i,i+m}^-}\Big\}\Bigg]^{\displaystyle{{\frac{1}{2}}}}.
\end{aligned}
\end{equation}
As the function of quantum coherence depend only on three terms of the density matrix (\ref{eq6}), this will make it easier to study its behavior numerically.
\section{Results}
\label{section4}
In  this  section,  we  present  our  numerical treatment based on the analytical approach of the entanglement, quantum discord, classical correlations and quantum coherence between second nearest (2N), third nearest (3N) and fourth nearest (4N) neighbors.
\subsection{Entanglement and Quantum Discord}
We start by comparing the behavior of the 2N, 3N and 4N spin pairs entanglement and quantum discord when the magnetic field is switched on, at $T=0$.
    \begin{figure}[H]
     \subfloat[Entanglement at zero temperature\label{fig4.1a}]{%
       \includegraphics[width=0.5\textwidth]{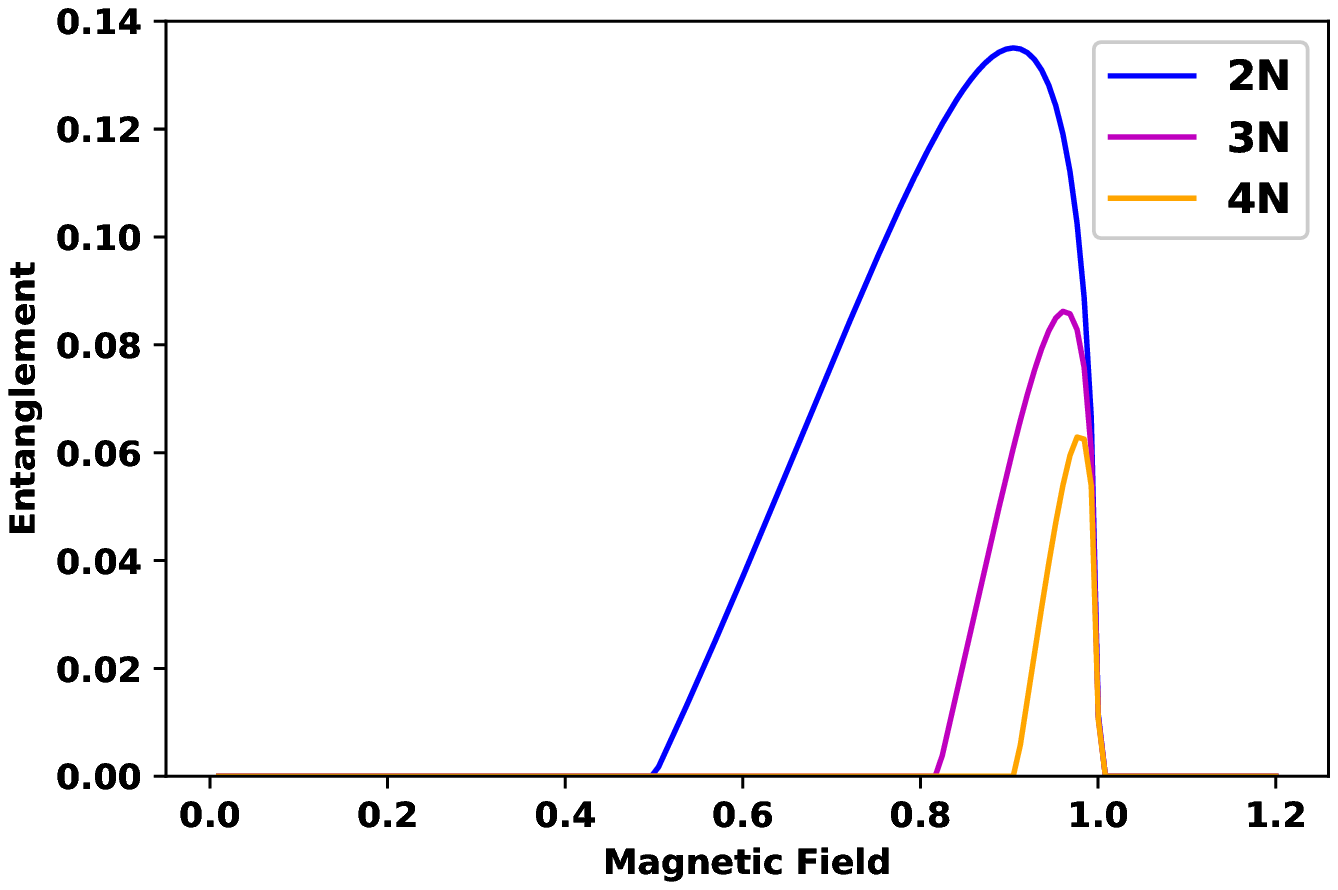}
       }
     \hfill
     \subfloat[Quantum discord at zero temperature\label{fig4.1b}]{%
       \includegraphics[width=0.5\textwidth]{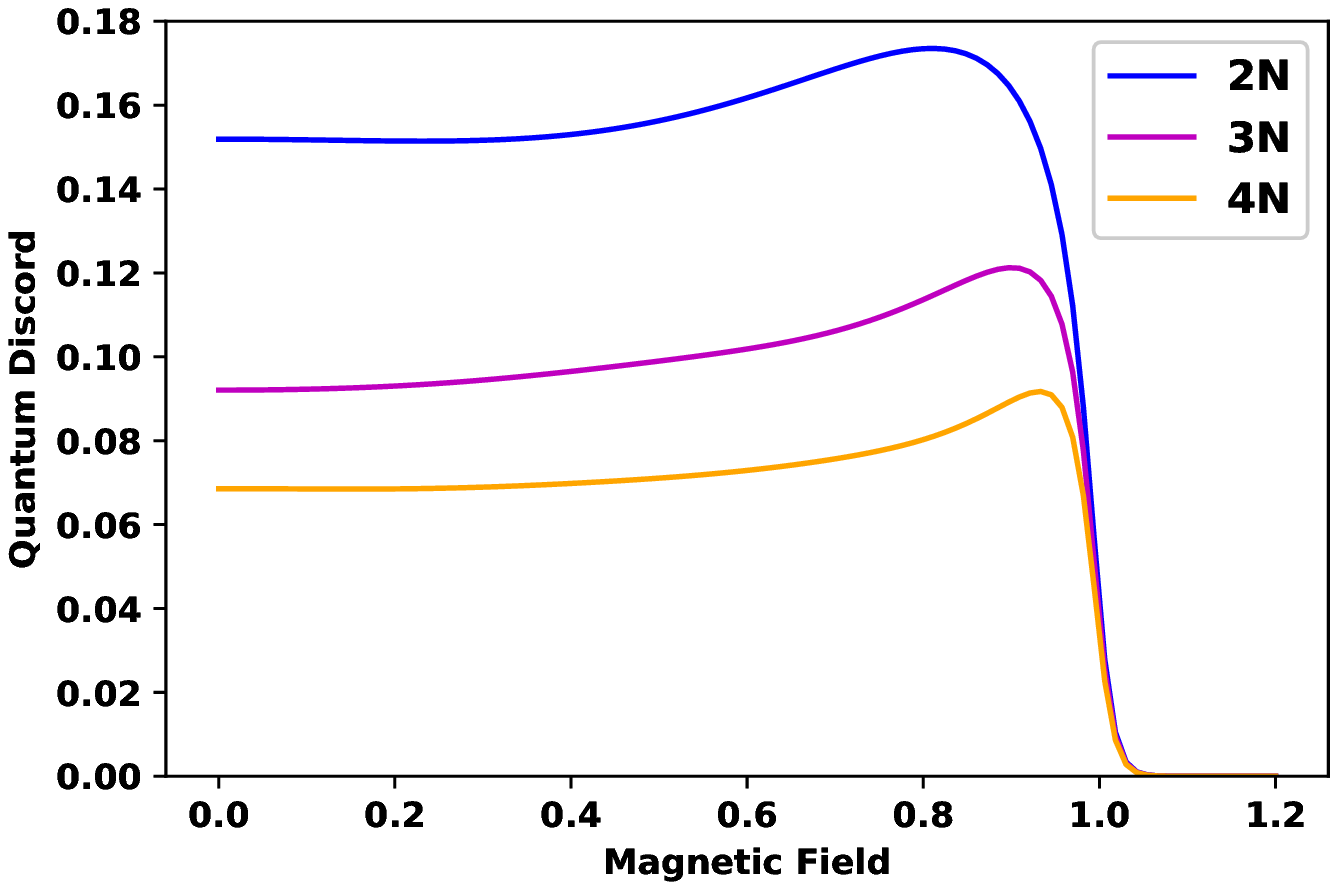}
       }
     \caption{The effect of the magnetic field on (a) entanglement and (b) quantum discord at $T=0$, for the 2N, 3N and 4N spin pairs. }
     \label{fig4.1}
   \end{figure}
Figure. \ref{fig4.1a} shows that when the magnetic field ``$h$" is switched off none of the spin pairs are entangled, and as we gradually increase the magnetic field we see creation of entanglement when $h$ reaches a critical value $h_c^E$ of $\frac{1}{2}$ for $2N$, $\frac{4}{5}$ for $3N$ and $\frac{9}{10}$ for $4N$.
Increasing the magnetic field beyond the critical value $h_c^E$, enhances entanglement between the spin pairs until its saturation at a critical value close to $h_d^E=1$ where it dies out and drops rapidely to zero in all three cases. In \cite{magnetic_ent} a relation between the critical field $h_c^E$ for entanglement creation and the critical field $h_d^E$ at which entanglement dies was proposed. A relation that is in agreement with the results from the finite size systems \cite{ent2}. For quantum discord however, the comportment is different. In Figure. \ref{fig4.1b} we see that quantum discord persists even when $h=0$ and at regions where entanglement is zero (i.e $h<h_d^E$). In general, quantum discord originates from the coherence that arises from quantum superposition, which exists in the subsystems of a quantum system and persists even if the system is in a product state, while
the amount of entanglement in a seperable state is always zero.
\begin{figure}[h]
    \centering
    \includegraphics[scale=0.5]{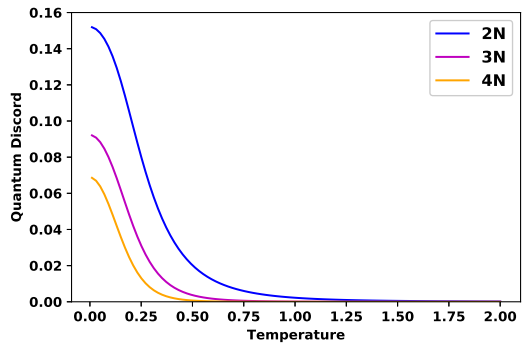}
    \caption{Quantum discord versus temperature at $h=0$, for the 2N, 3N and 4N spin pairs.}
    \label{fig4.2}
\end{figure}

Next, we analyze the effect of the temperature on entanglement and quantum discord when the magnetic field is switched off. As depicted in Figure. \ref{fig4.2}, in all three cases, quantum discord starts with a maximum value and decreases as $T$ increases, and tends asymptotically to zero. This is in contrast to the case of entanglement where from Figure. \ref{fig4.1a} it is clear that entanglement is absent when $h=0$.
This behavior is expected as when one increases the temperature, the quantum fluctuations are dominated by the thermal ones, but this does not kill the quantumness of the system. The process that make it vanish is decoherence, and since absence of entanglement does not imply classicality and the fact that quantum discord is more robust under decoherence, explains the asymptotic behavior of the total quantum correlations as dipected in Figure. \ref{fig4.2}. This shows that product states are not negligible at all \cite{qc_persist} and it also confirms the robustness of quantum discord against temperature and its potential usefulness as resource for quantum technologies.
    \begin{figure}[H]
     \subfloat[2N\label{fig4.4a}]{%
       \includegraphics[width=0.32\textwidth]{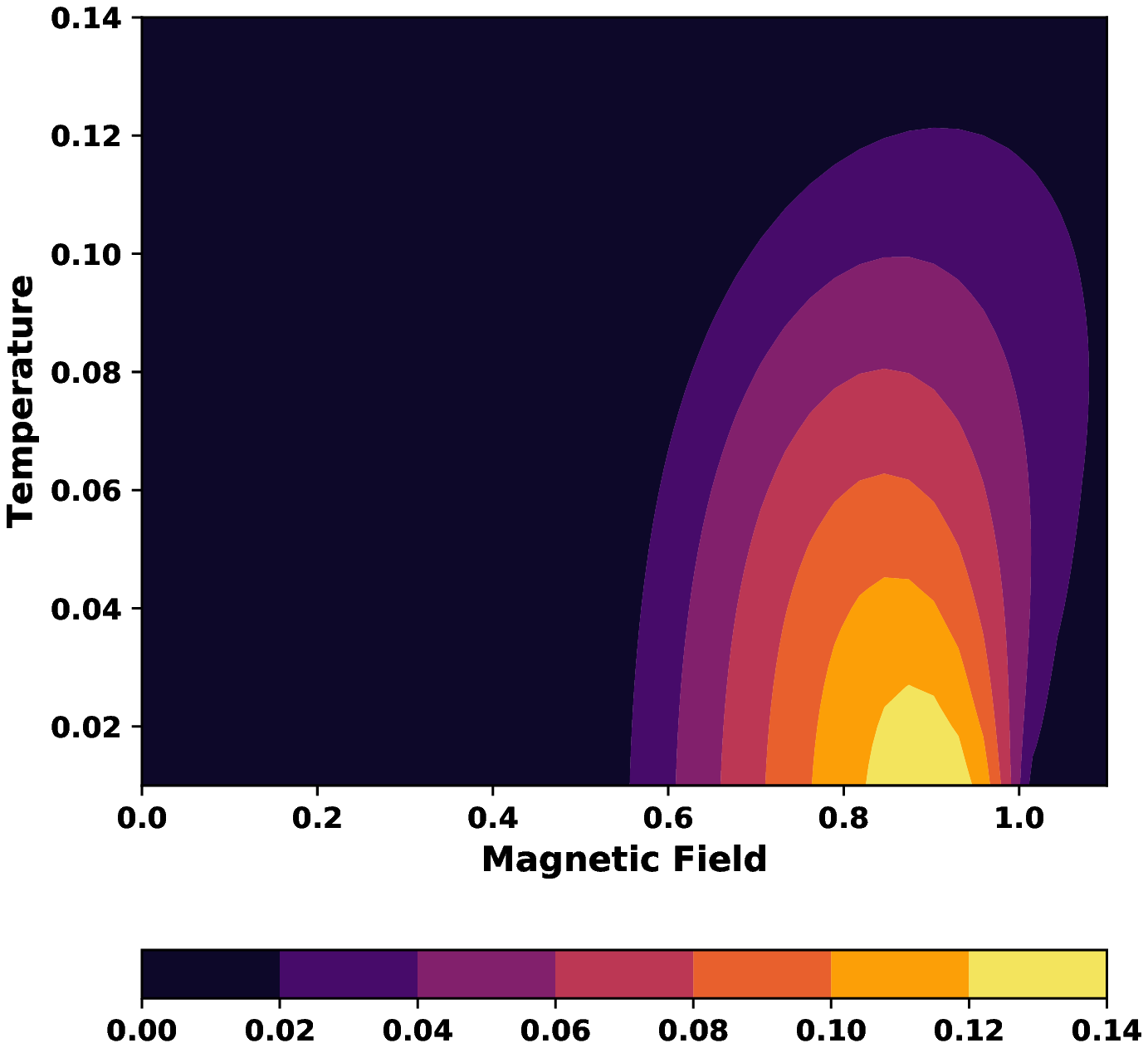}
       }
     \hfill
     \subfloat[3N\label{fig4.4b}]{%
       \includegraphics[width=0.32\textwidth]{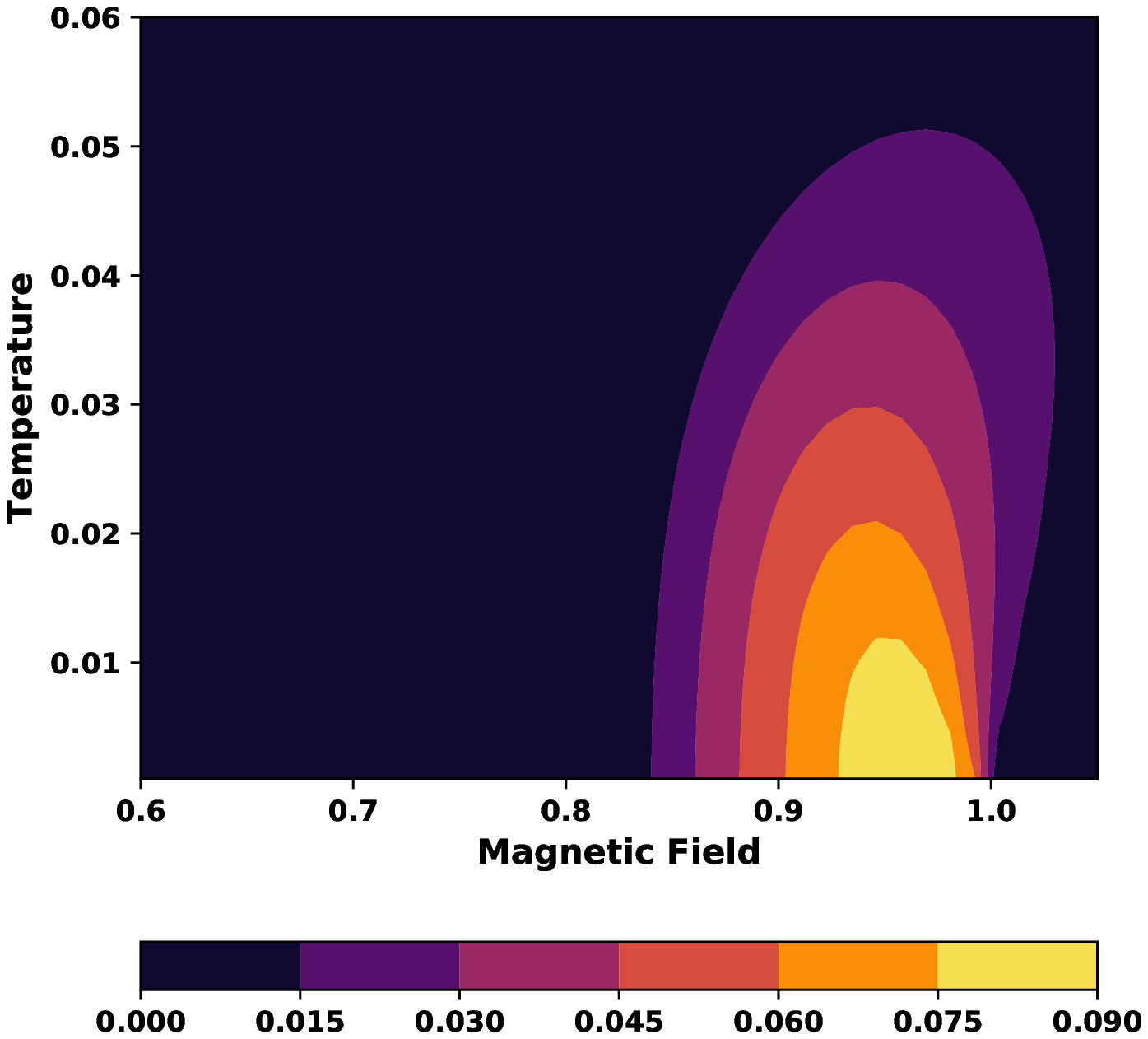}
       }
     \hfill
     \subfloat[4N\label{fig4.4c}]{%
       \includegraphics[width=0.32\textwidth]{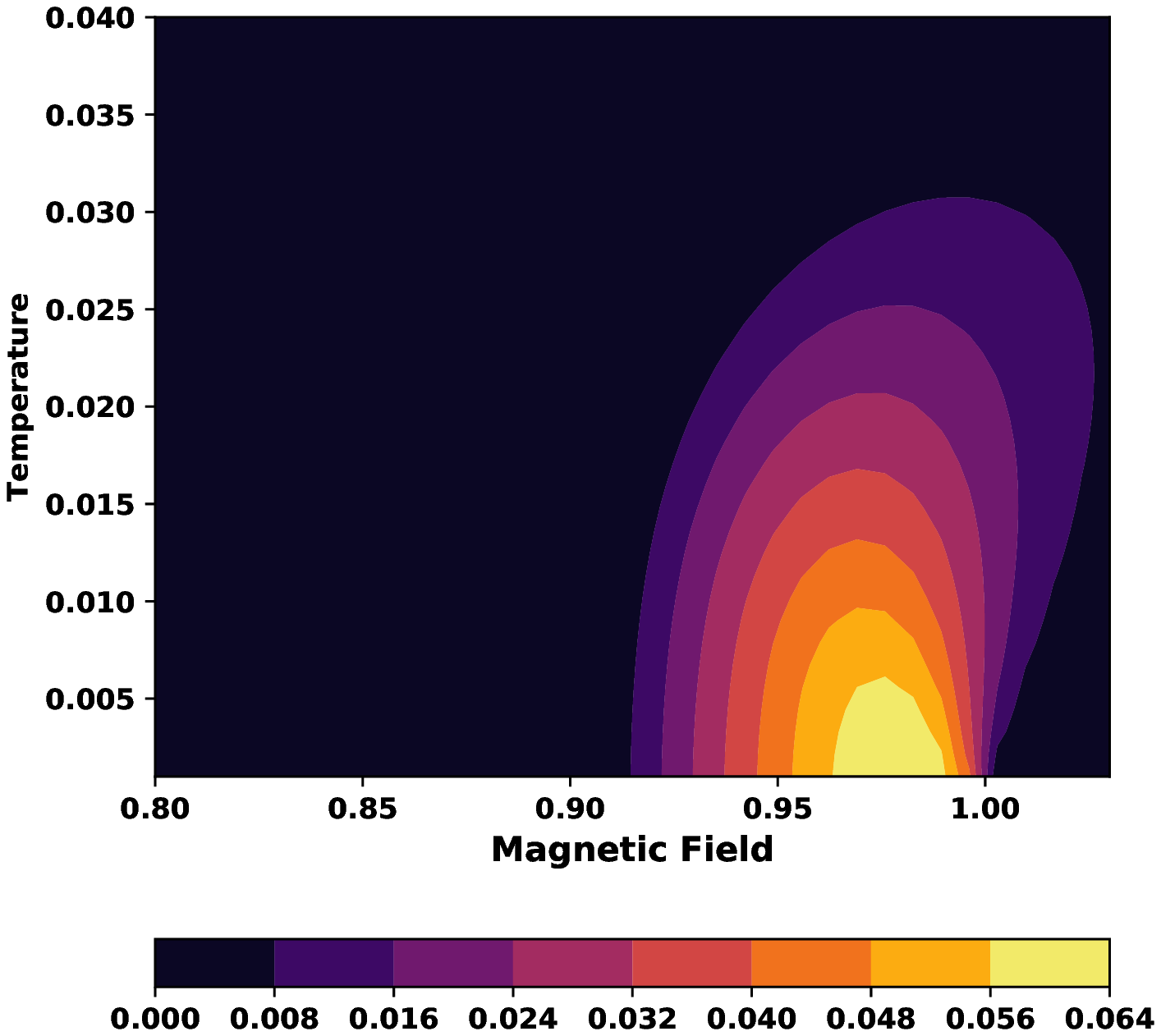}
       }
     \caption{$T-h$ Phase diagram of entanglement for the 2N, 3N and 4N spin pairs.}
     \label{fig4.4}
   \end{figure}
The $T-h$ phase diagram of the entanglement and quantum discord of the 2N, 3N, 4N spin pairs is depicted in Figure. \ref{fig4.4} and Figure. \ref{fig4.5}. It is clear from the figures that as the distance between the pairs increases, the quantum correlations magnitude decreases which is expected in the sense that as the distance increase the strength of the correlations decreases. Furthermore, from Figure. \ref{fig4.4} entanglement is seen to be easily destroyed by the temperature as the distance between pairs increase, while quantum discord is more robust as shown in Fig. \ref{fig4.5}. Such behavior is due to the dominance of thermal fluctuation over quantum correlations when one increases the temperature. However, the thermal and magnetic interval of entanglement in the 3N and 4N spin pairs is smaller than the 2N pairs, while for quantum discord this does not change, it is defined in the entire interval.
    \begin{figure}[H]
     \subfloat[2N\label{fig4.5a}]{%
       \includegraphics[width=0.32\textwidth]{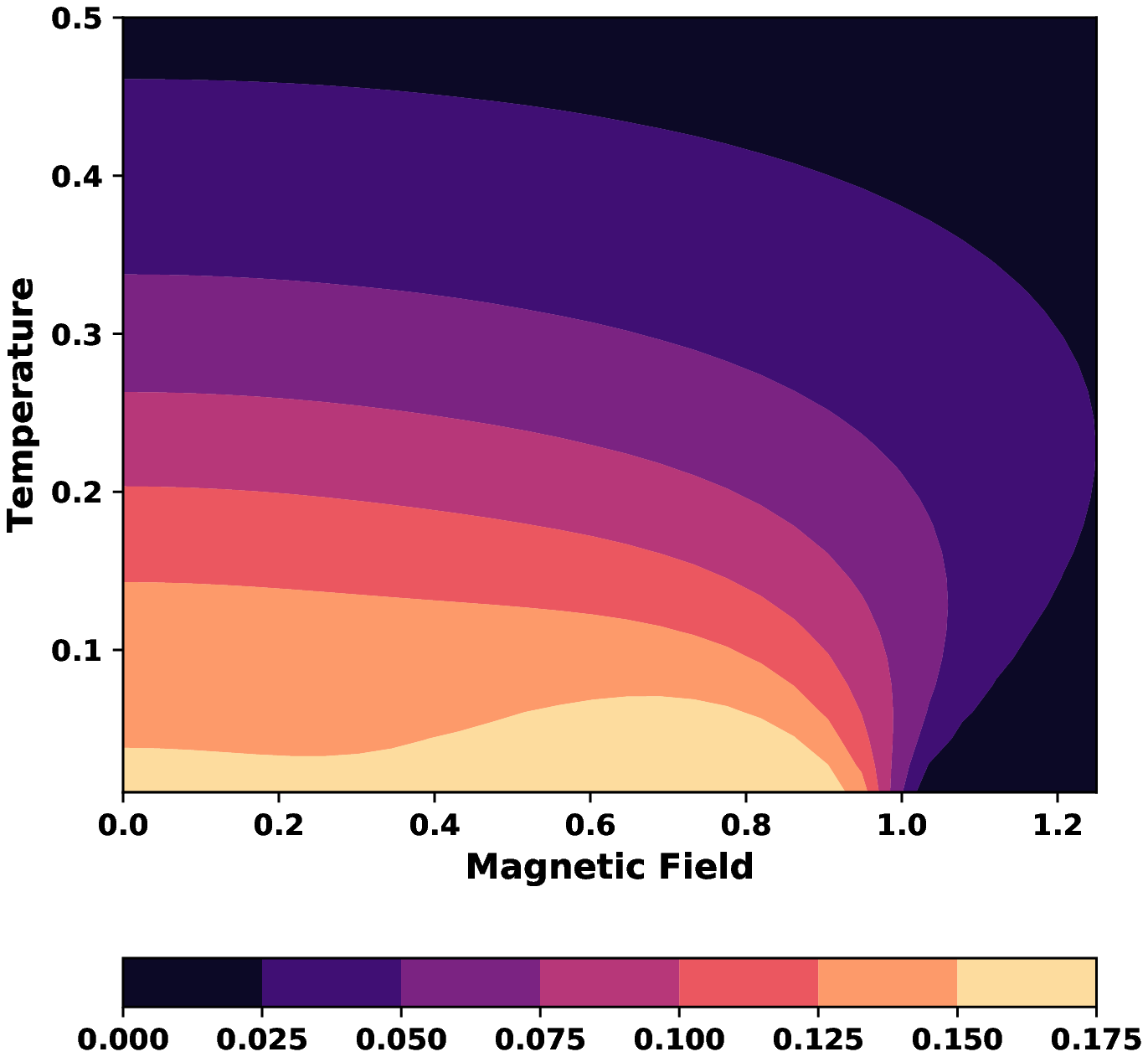}
       }
     \hfill
     \subfloat[3N\label{fig4.5b}]{%
       \includegraphics[width=0.32\textwidth]{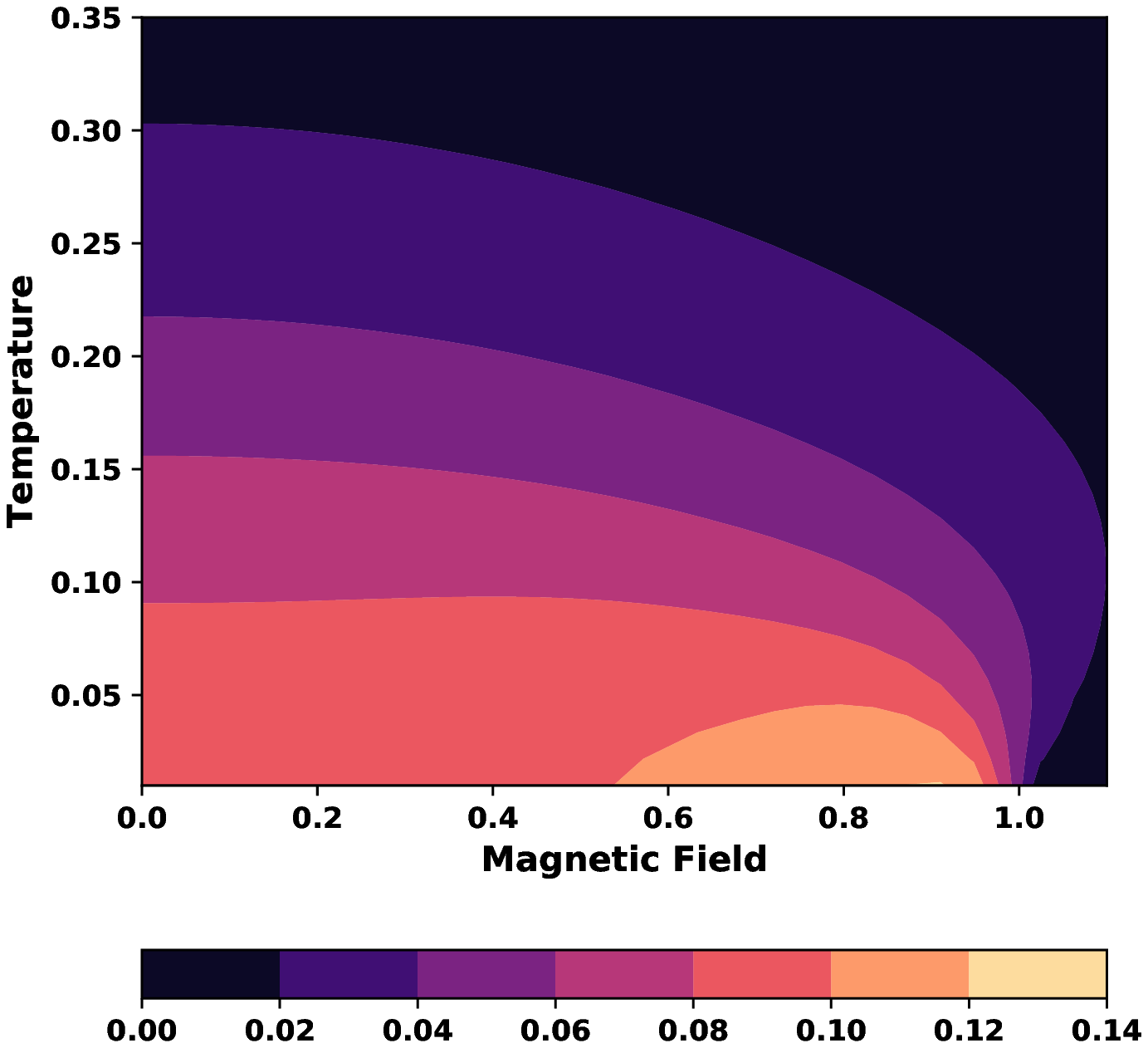}
       }
     \hfill
     \subfloat[4N\label{fig4.5c}]{%
       \includegraphics[width=0.32\textwidth]{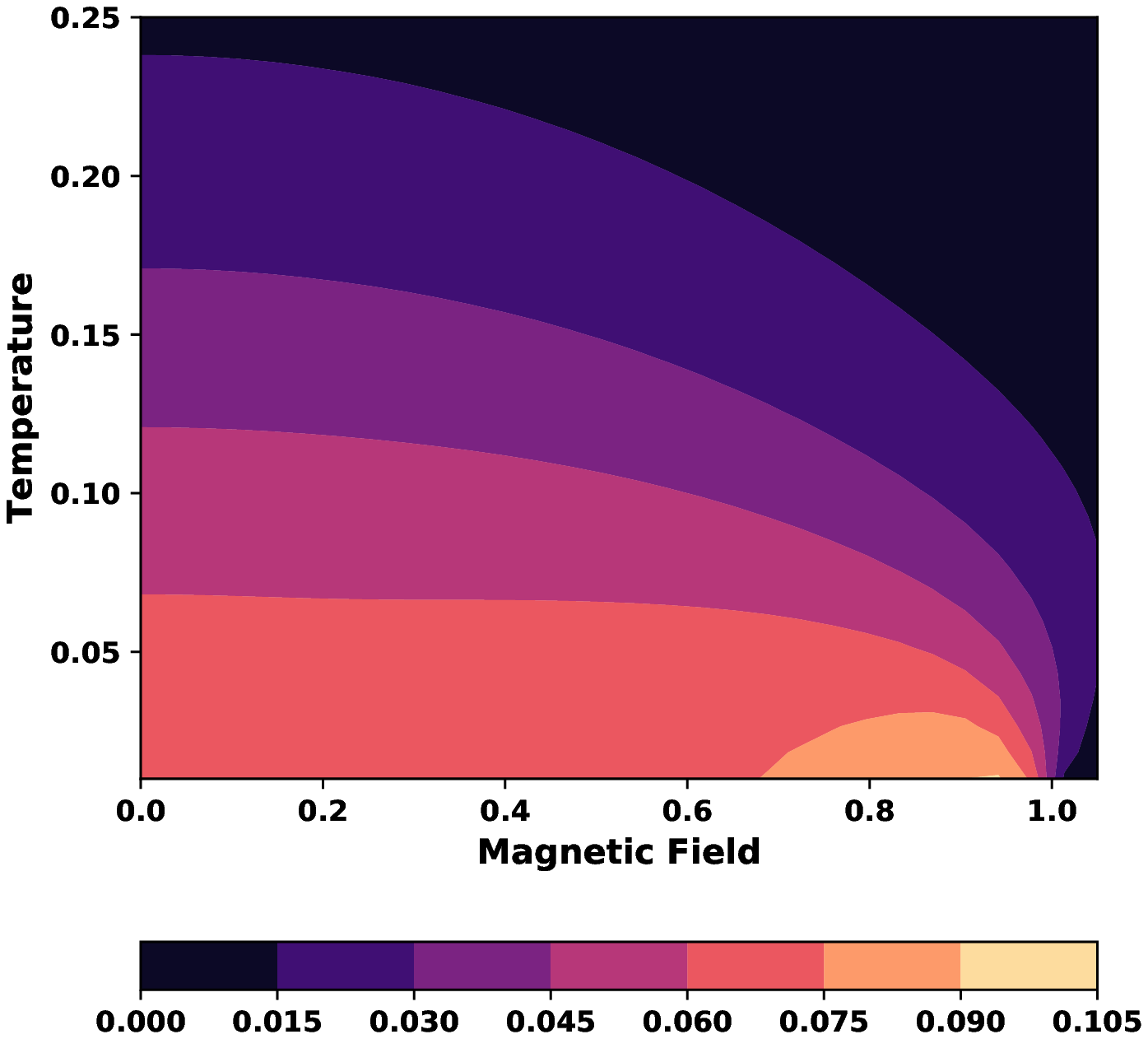}
       }
     \caption{$T-h$ Phase diagram of quantum discord, for the 2N, 3N and 4N spin pairs. }
     \label{fig4.5}
   \end{figure}

\subsection{Classical Correlations}
\begin{figure}[H]
     \subfloat[2N\label{fig4.8a}]{%
       \includegraphics[width=0.32\textwidth]{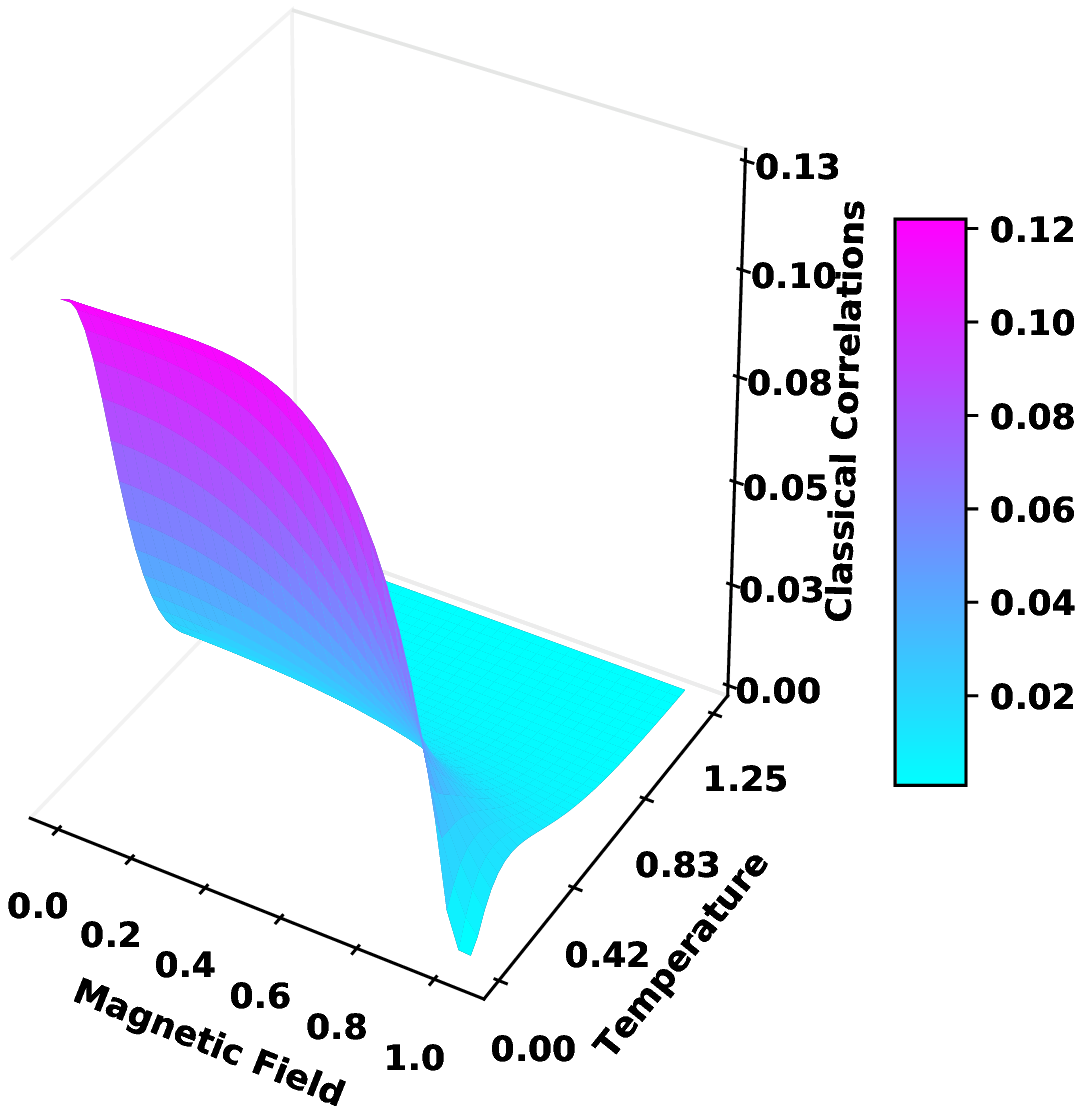}
       }
     \hfill
     \subfloat[3N\label{fig4.8b}]{%
       \includegraphics[width=0.32\textwidth]{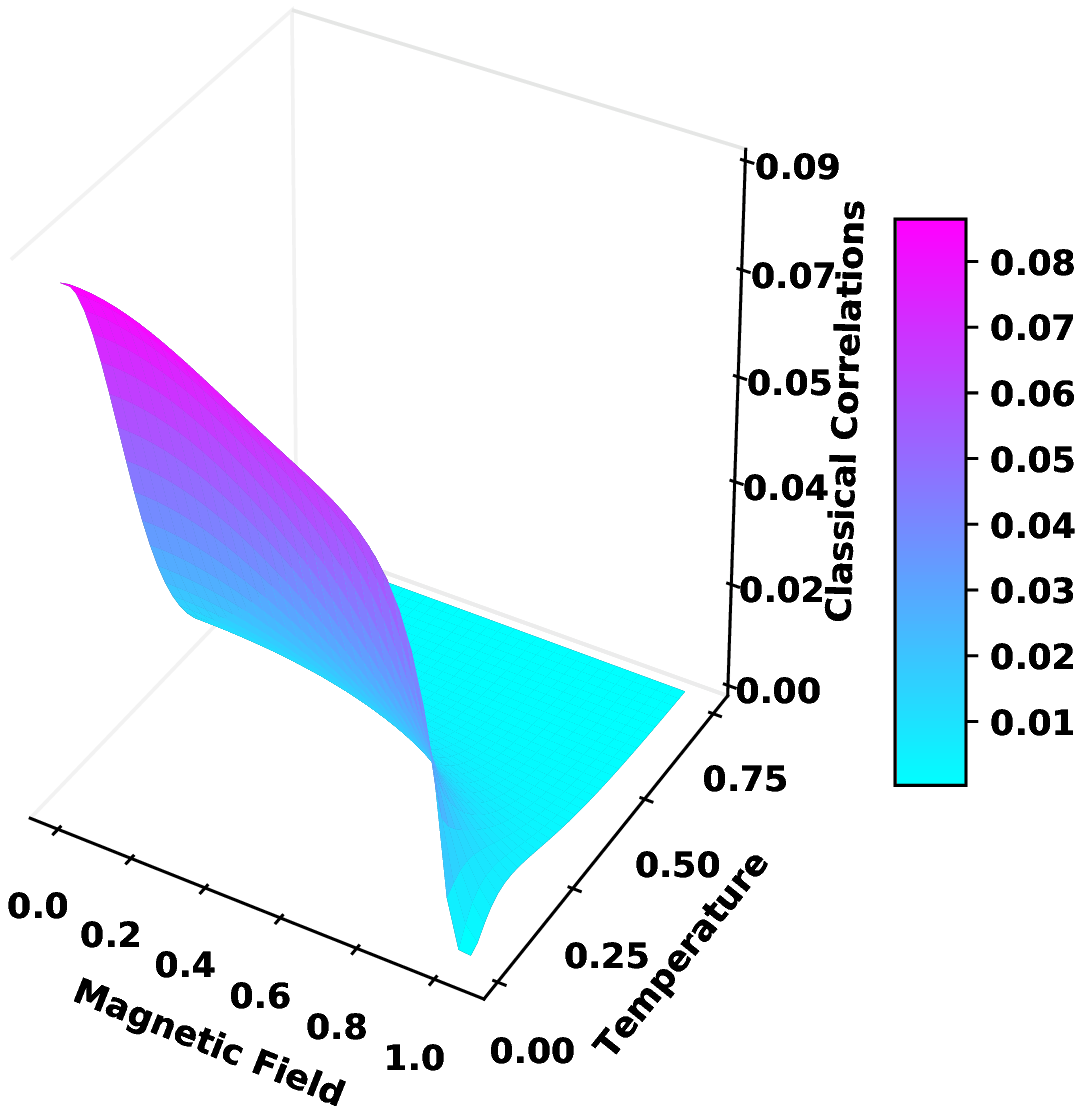}
       }
     \hfill
     \subfloat[4N\label{fig4.8c}]{%
       \includegraphics[width=0.32\textwidth]{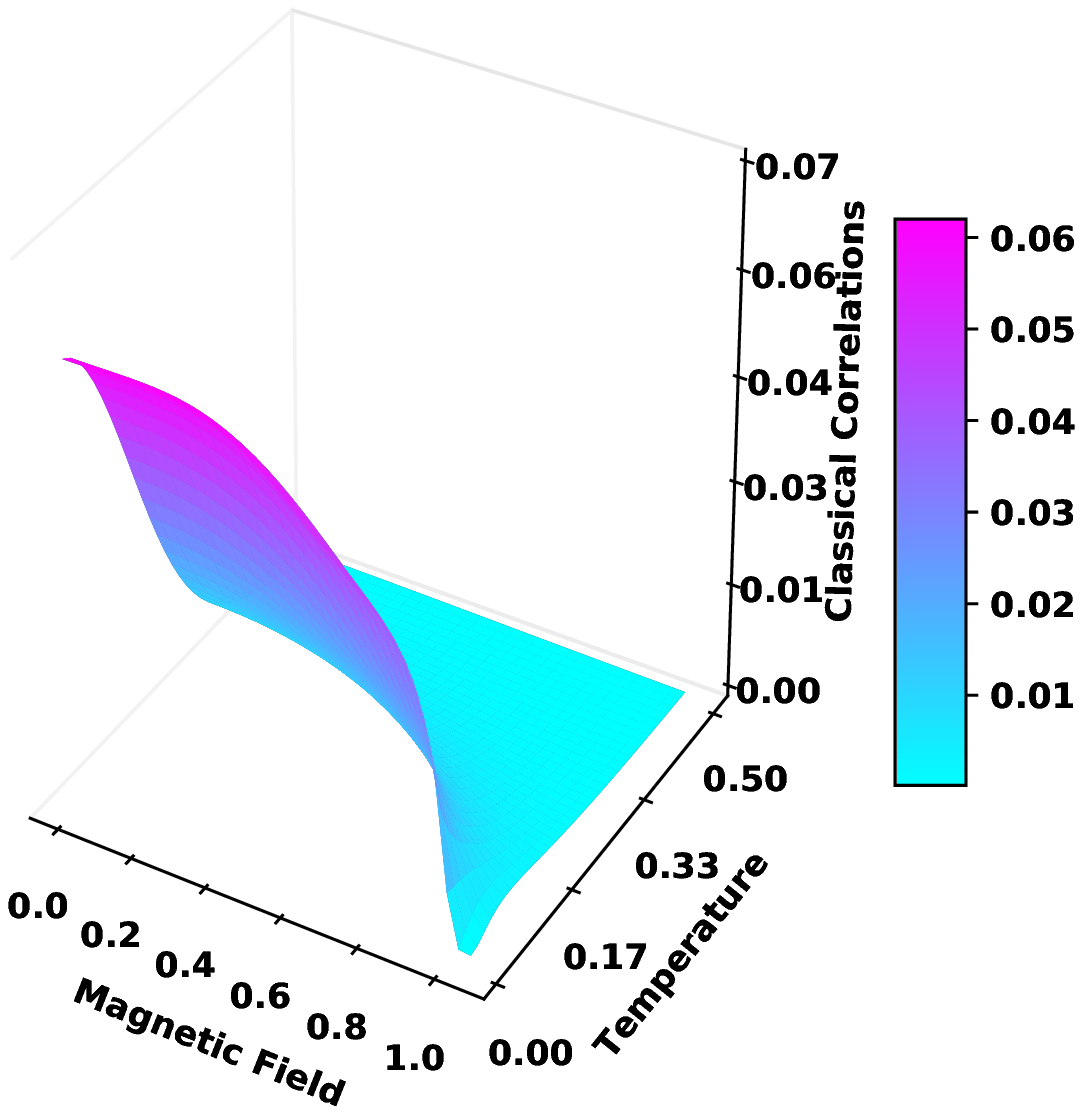}
       }
     \caption{3D panorama of classical correlations for the 2N, 3N and 4N spin pairs}
     \label{fig4.6}
   \end{figure}

The behavior of classical correlations in the 1D XX chain is addressed in this subsection. We plot in Figure. \ref{fig4.6} a 3D panorama of classical correlations for the 2N, 3N and 4N spin pairs. As seen for entanglement and quantum discord in the previous subsection, classical correlations also decreases with distance between spins and classicality persists in the overall thermal and magnetic interval. Furthermore, in comparison with the behavior of quantum discord in Figure. \ref{fig4.5}, we see the dominance of quantum correlations over their classical counterpart while in the transverse Ising model the inverse phenomenon takes place \cite{discord_zero3}. More interestingly, as it can be seen from Figure. \ref{fig4.5} in the presence of the temperature quantum correlations can be increased in some regions by the presence of the magnetic field, while for classical correlations it decreases with respect to the magnetic field for zero and finite values of the temperature.
 
\subsection{Quantum Coherence}
\begin{figure}[H]
     \subfloat[Variation of Quantum Coherence with respect to the Temperature\label{fig4.7a}]{%
       \includegraphics[width=0.5\textwidth]{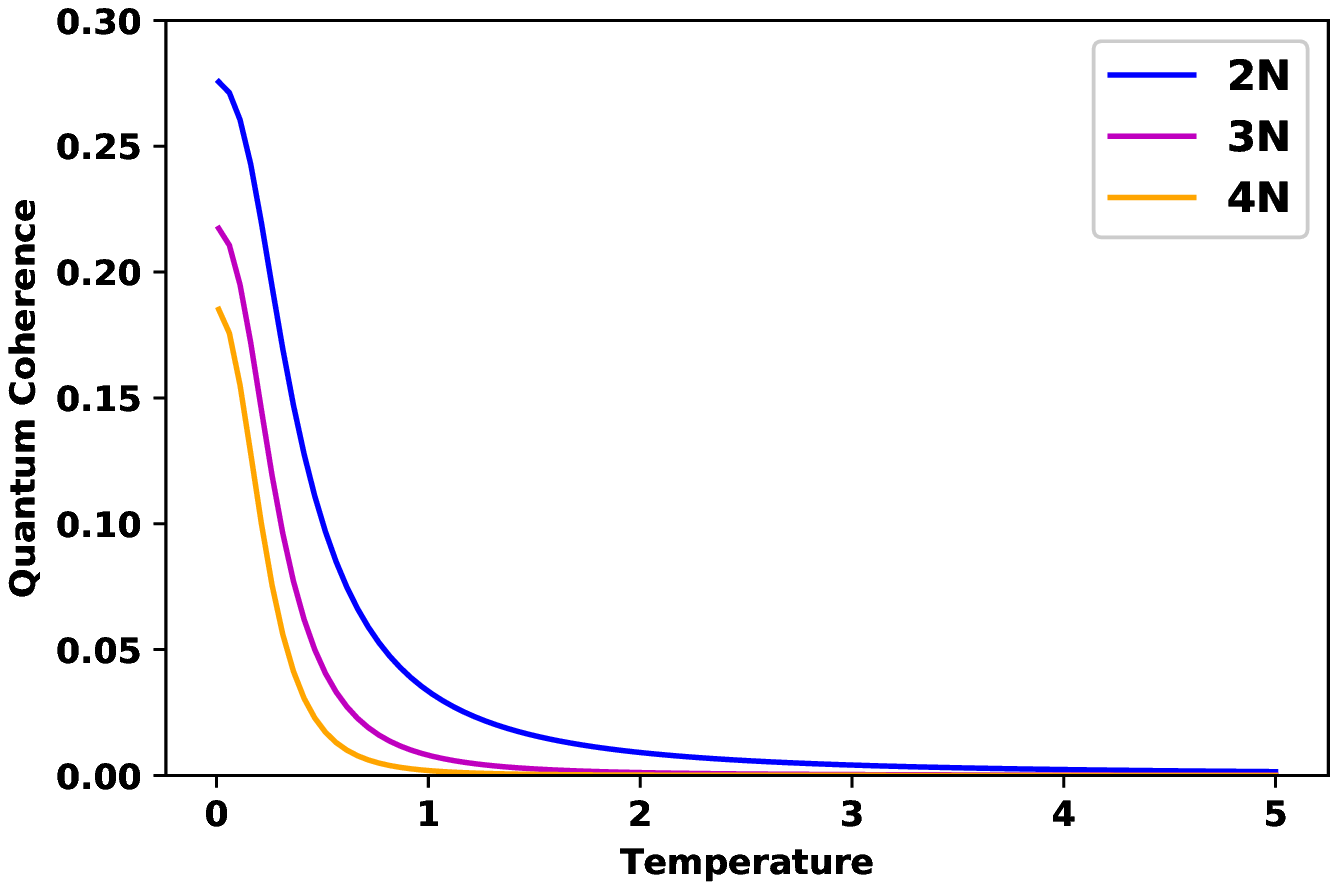}}
     \hfill
     \subfloat[Variation of Quantum Coherence with respect to the magnetic field\label{fig4.7b}]{%
       \includegraphics[width=0.5\textwidth]{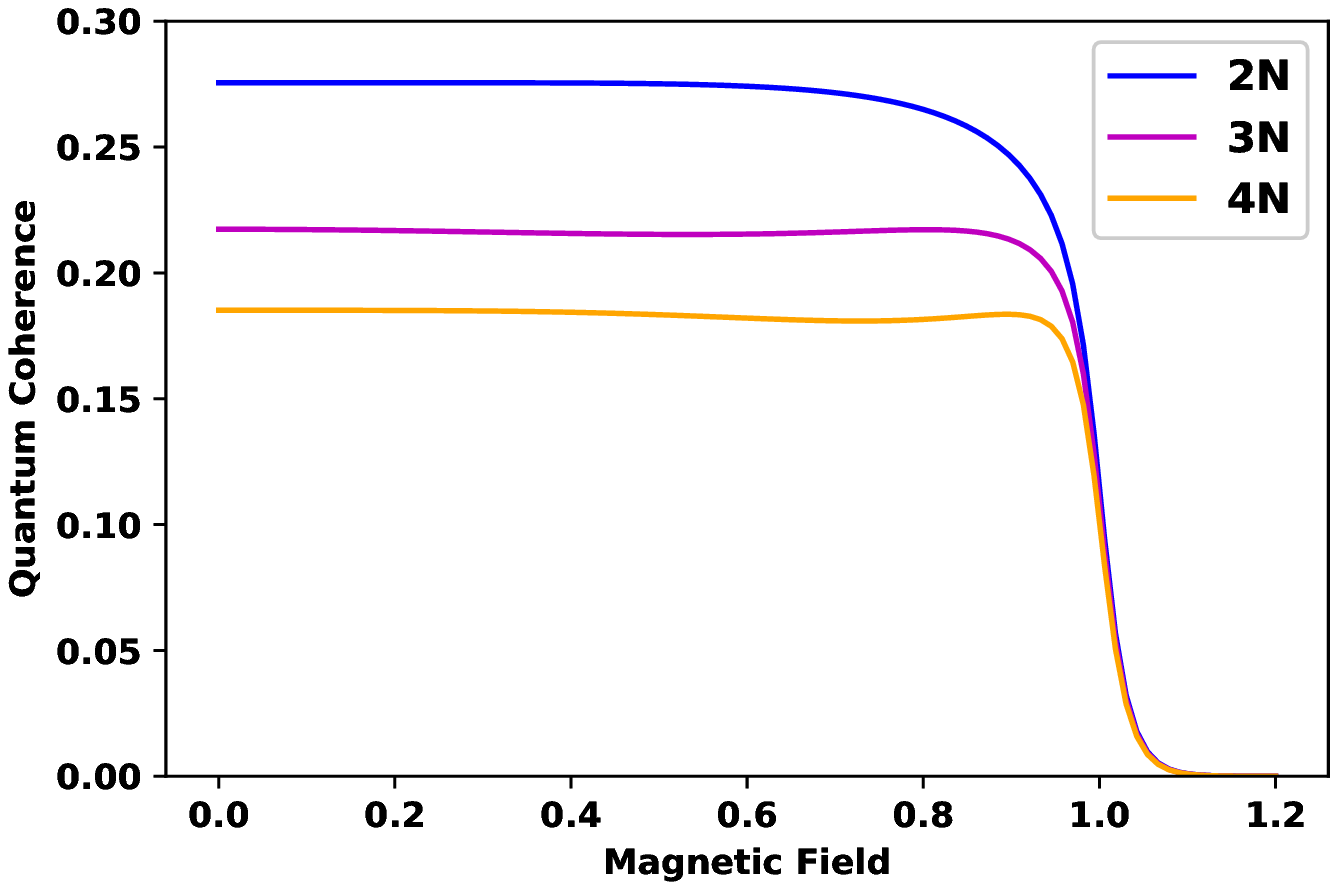}
       }
     \caption{Effect of (a) the temperature and (b) the magnetic field on Quantum Coherence for several lattice spacing}
     \label{fig4.7}
   \end{figure}

We conclude this section by a study of quantum coherence in the XX model. Figure. \ref{fig4.7} shows the typical behavior of quantum coherence with respect to the temperature and the magnetic field for the 2N, 3N and 4N spin pairs, in which we observe the similarity with quantum discord depicted previously in  Fig. \ref{fig4.1b} and Fig. \ref{fig4.2}; this can be explained as follows.

In general, coherence stems from individual sites (local coherence) or may be distributed along sites (intrinsic coherence) \cite{coherence_distri} and, one can show that due to spin-flip symmetry of the XX chain, the one site density matrix is diagonal which means that the local coherence as defined in \cite{coherence_distri} is zero for the XX model. So, the total coherence observed in the system will entirely originate from the correlations between the two spins, and since quantum discord represents total quantum correlations, the behavior of quantum coherence will be similar to that of quantum discord. It is worthwhile noting that in some models (e.g. Heisenberg spin models with Dzyaloshinsky-Moriya interactions \cite{sp_coherence}) the inverse behavior is observed, and the behavior of quantum coherence is reminiscent of entanglement not quantum discord.\\Moreover, in \cite{discrepancy} it was found that quantum discord can be understood from the discrepancy between the relative quantum coherence for the total system and that of the subsystem chosen for the calculation of quantum discord. This set up a clear interplay between quantum correlations and quantum coherence.

\begin{figure}[H]
     \subfloat[2N\label{fig4.8a}]{%
       \includegraphics[width=0.32\textwidth]{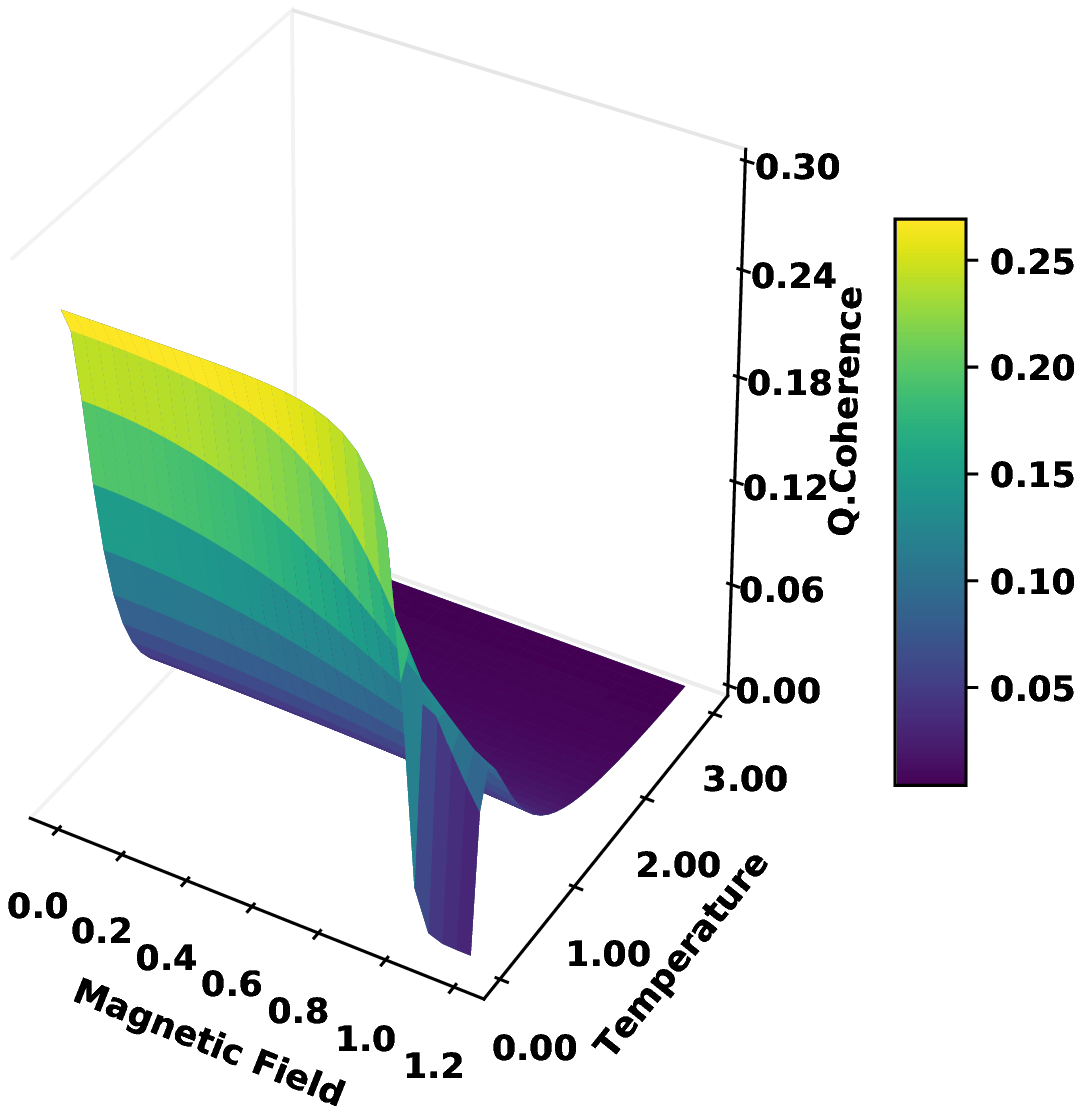}
       }
     \hfill
     \subfloat[3N\label{fig4.8b}]{%
       \includegraphics[width=0.32\textwidth]{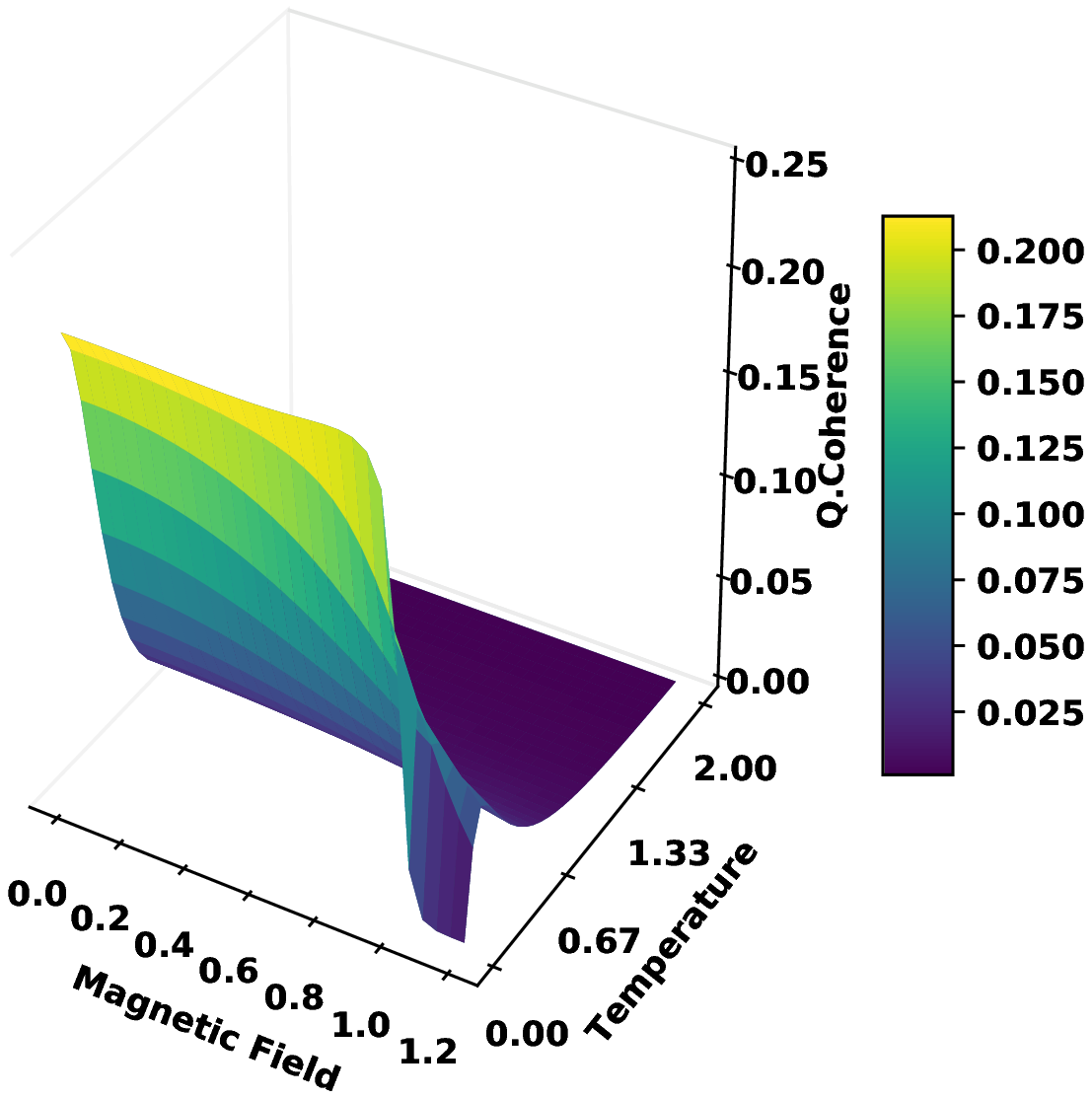}
       }
     \hfill
     \subfloat[4N\label{fig4.8c}]{%
       \includegraphics[width=0.32\textwidth]{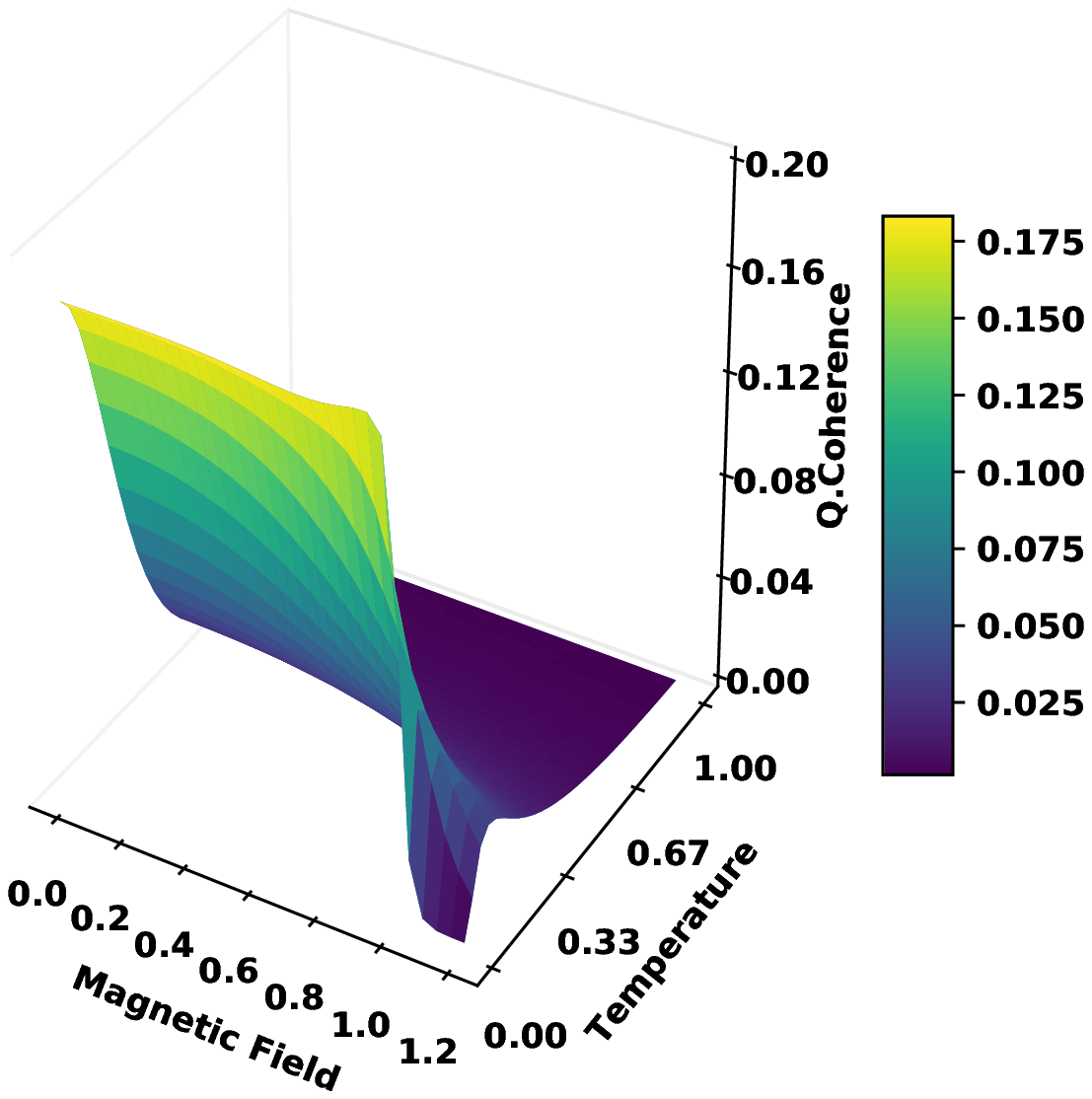}
       }
     \caption{3D plot of quantum coherence for the 2N, 3N and 4N spin pairs}
     \label{fig4.8}
    \end{figure}

Finally, a representation of the complete spectrum of quantum coherence for the 2N, 3N and 4N spin pairs is sketched in Figure. \ref{fig4.8}, in which we see the long-range property of quantum coherence. 
\noindent This characteristic makes it a candidate, like quantum discord, to characterize quantum phase transitions \cite{coherence_spin4}. We also observe from the figure that like the other types of correlations studied in this paper the magnitude of quantum coherence decreases with the distance between the spins.\\ 
Furthermore, from Fig. \ref{fig4.7a} and Fig. \ref{fig4.8} we see that quantum coherence decreases with the temperature which is a similar treat with respect to other types of correlations studied in this paper. For quantum correlations, this is due in general to the  dominance of thermal fluctuation over quantum fluctuations, when we reach a threshold temperature as we increase it, which forces quantum correlations to decay. However, what is more interesting about quantum coherence is that it outperforms entanglement and quantum discord in the robustness to the temperature as it can be seen from Figure. \ref{fig4.7a} and Figure. \ref{fig4.8} which is a direct consequence of the inclusion relation : Entanglement $\subset$ Quantum discord $\subset$ Quantum coherence. Knowing that entanglement is a form of coherence and given the non-local nature of quantum coherence, this may make it an outstanding candidate from the perspective of resource theory to perform quantum information processing task using quantum spin chains.
\subsection{Quantum Phase Transitions}
\begin{figure}[t]
     \subfloat[Quantum entanglement (QE)\label{fig4.9a}]{%
       \includegraphics[width=0.32\textwidth]{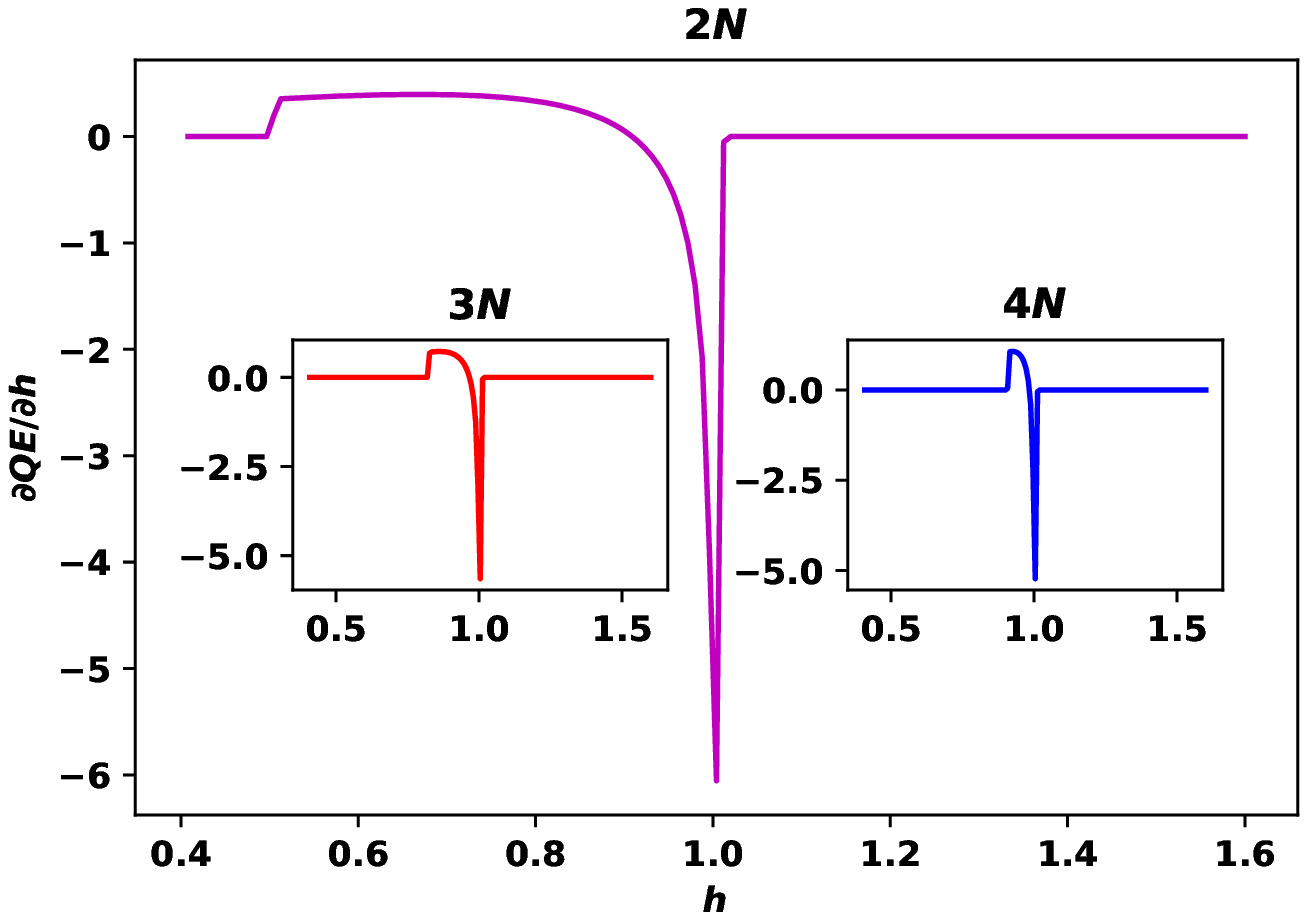}
       }
     \hfill
     \subfloat[Quantum discord (QD)\label{fig4.9b}]{%
       \includegraphics[width=0.32\textwidth]{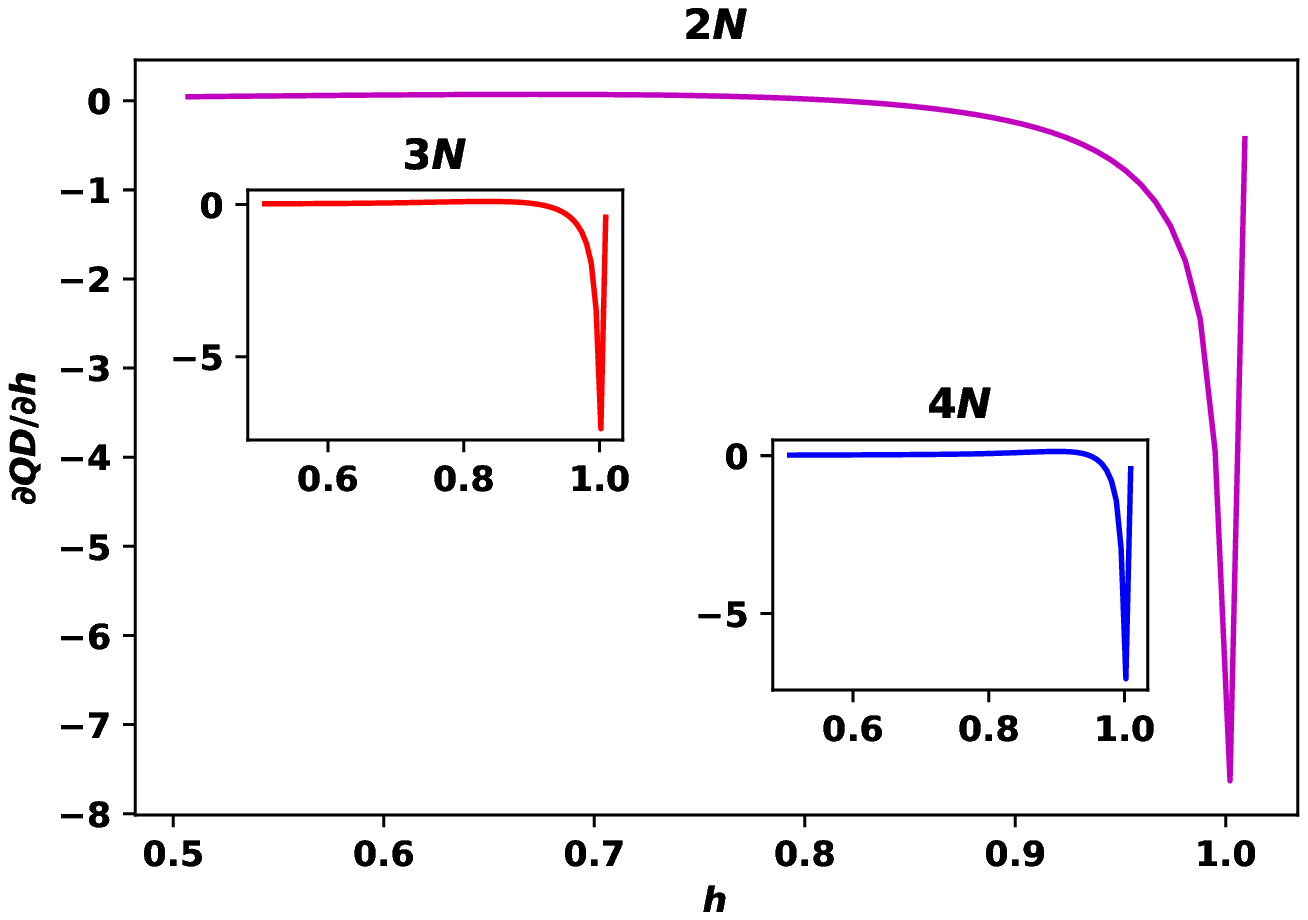}
       }
     \hfill
     \subfloat[Quantum coherence (QC)\label{fig4.9c}]{%
       \includegraphics[width=0.32\textwidth]{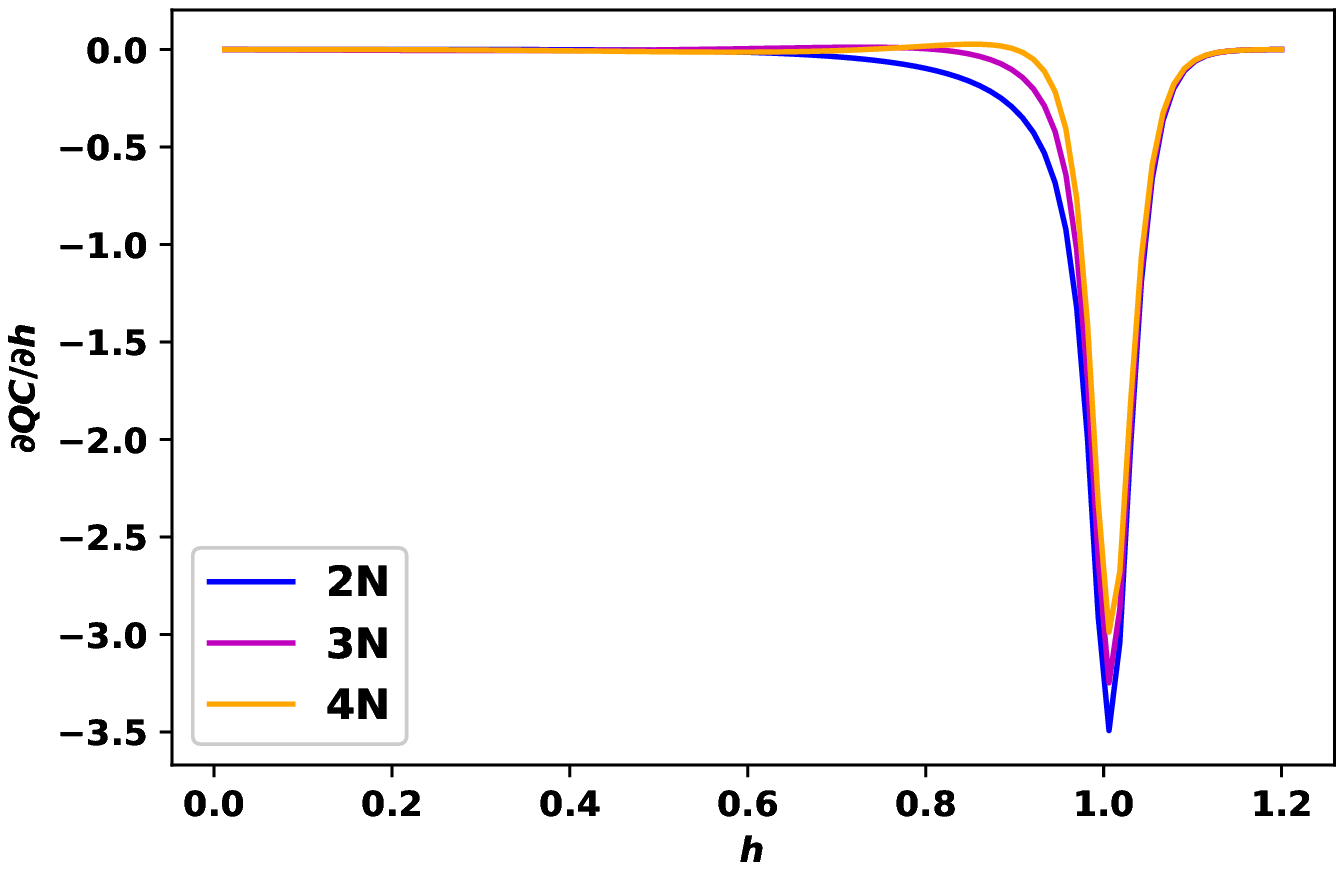}
       }
     \caption{First derivative of entanglement, quantum discord and quantum coherence with respect to the magnetic field $h$, for the 2N, 3N and 4N spin pairs}
     \label{fig4.9}
    \end{figure}

The XX model is known to exhibit quantum phase transitions (QPT). At the thermodynamic limit, which is the case studied here, the XX model undergoes only one continuous QPT at a critical field $h_c=J=1$, where the system transits from Mott-insulator phase to superfluid phase or vice versa. The ground state of the Mott-insulator phase is a separable state while the ground state of superfluid phase is sure to be an entangled state \cite{sachdev,qpt_xx}. In general, QPT are due to quantum fluctuations which stems from the Heisenberg uncertainty principle \cite{sachdev}, and these fluctuations can be captured by quantum correlations. Moreover, QPT is followed by a specific change in the nature of quantum correlations in the ground state of the system and this line of thoughts is what led at the beginning to the investigations of QPT in the vicinity of entanglement \cite{ent_nielsen}.\\
Studying the derivative of entanglement with respect to the magnetic field reveals the quantum critical point clearly. In Fig. \ref{fig4.9a}, the derivative of quantum entanglement shows a singular behavior for the 2N, 3N and 4N spin pairs, at $h_c=1$ which is the critical point of the model associated with the quantum phase transition. Which is in agreement with studies in other models.
\\However, relying on entanglement which is known to be a short-ranged quantity has been proven inefficient in other models \cite{discord_finite1,qpt_xy}, like in the case of the Ising model, where the entanglement vanishes for sites farther than 2N spin neighbors \cite{discord_zero3}. For this purpose, alternatives like multipartite entanglement, quantum discord and quantum coherence are needed in the study of QPT in more general models.\\
Finally, we conclude this section by studying the behavior of the derivative of quantum discord and quantum coherence with respect to the magnetic field, where in Fig. \ref{fig4.9b} and Fig. \ref{fig4.9c} we see that both measures show a singular behavior like entanglement at the quantum critical point $h_c=1$ for the 2N, 3N and 4N spin pairs, meaning that they also reveal the quantum phase transition in the XX model.
\section{Conclusion}
In summary, we presented in this paper a comparative study between various types of pairwise correlations that may emerge in a quantum spin system, such as : entanglement, quantum discord, classical correlations and quantum coherence. The analytical expressions of the pairwise quantities were provided using the Jordan-Wigner transformation which we also verified numerically, and the ``Pauli basis expansion". We then, showed the dominance of quantum correlations over their classical counterpart, the potential use of quantum discord in quantum information processing being robust against the temperature, and the role of quantum entanglement, quantum discord and quantum coherence in detecting QPT in the XX model. Finally, we focused on quantum coherence, an essential feature of quantum mechanics which we proved to be a long range quantity and outperforming entanglement and quantum discord in the robustness to temperature. This motivates the generalization of the study of quantum coherence to more generalized models in higher dimensions, and also to perform quantum information processing using quantum spin chains, with quantum coherence as a resource.
\label{section5}
\section*{Acknowledgments}
It is our pleasure to thank Prof. Abderrahmane Maaouni for his helpful advices on various numerical issues encountered in this paper. Zakaria Mzaouali would like to thank ICTP-Trieste for its hospitality during his visit, where part of the work was done.



\bibliographystyle{elsarticle-num}

\bibliography{sample}

\begin{thebibliography}{10}
\expandafter\ifx\csname url\endcsname\relax
  \def\url#1{\texttt{#1}}\fi
\expandafter\ifx\csname urlprefix\endcsname\relax\def\urlprefix{URL }\fi
\expandafter\ifx\csname href\endcsname\relax
  \def\href#1#2{#2} \def\path#1{#1}\fi

\bibitem{schrodinger1935}
E.~Schr\"odinger, Discussion of probability relations between separated
  systems, Math. Proc. Cambridge Philos. Soc. 31~(4) (1935) 555–563.

\bibitem{entanglement_review}
R.~Horodecki, P.~Horodecki, M.~Horodecki, K.~Horodecki, Quantum entanglement,
  Rev. Mod. Phys. 81 (2009) 865--942.

\bibitem{epr}
A.~Einstein, B.~Podolsky, N.~Rosen, Can quantum-mechanical description of
  physical reality be considered complete?, Phys. Rev. 47 (1935) 777--780.

\bibitem{teleportation}
C.~H. Bennett, G.~Brassard, C.~Cr\'epeau, R.~Jozsa, A.~Peres, W.~K. Wootters,
  Teleporting an unknown quantum state via dual classical and
  einstein-podolsky-rosen channels, Phys. Rev. Lett. 70 (1993) 1895--1899.

\bibitem{qcomunication}
H.~Salih, Z.-H. Li, M.~Al-Amri, M.~S. Zubairy, Protocol for direct
  counterfactual quantum communication, Phys. Rev. Lett. 110 (2013) 170502.

\bibitem{qcomputation}
C.~H. Bennett, D.~P. DiVincenzo, Quantum information and computation, Nature
  404 (2000) 247 EP.

\bibitem{rosario}
L.~Amico, R.~Fazio, A.~Osterloh, V.~Vedral, Entanglement in many-body systems,
  Rev. Mod. Phys. 80 (2008) 517--576.

\bibitem{ent_nielsen}
T.~J. Osborne, M.~A. Nielsen, Entanglement in a simple quantum phase
  transition, Phys. Rev. A 66 (2002) 032110.

\bibitem{quantum_magnetism}
S.~Ulrich, R.~Johannes, F.~Damian~J.J., B.~Raymond~F., Quantum Magnetism, 645,
  Springer-Verlag Berlin Heidelberg, 2004.

\bibitem{creation_qsc_1}
L.-M. Duan, E.~Demler, M.~D. Lukin, Controlling spin exchange interactions of
  ultracold atoms in optical lattices, Phys. Rev. Lett. 91 (2003) 090402.

\bibitem{creation_qsc_2}
I.~Buluta, F.~Nori, Quantum simulators, Science 326~(5949) (2009) 108--111.

\bibitem{discord_vedral}
L.~Henderson, V.~Vedral, Classical, quantum and total correlations, J. Phys. A
  : Math. Gen. 34~(35) (2001) 6899.

\bibitem{quantum_correlation}
J.~Oppenheim, M.~Horodecki, P.~Horodecki, R.~Horodecki, Thermodynamical
  approach to quantifying quantum correlations, Phys. Rev. Lett. 89 (2002)
  180402.

\bibitem{discord}
H.~Ollivier, W.~H. Zurek, Quantum discord: A measure of the quantumness of
  correlations, Phys. Rev. Lett. 88 (2001) 017901.

\bibitem{discord_zero1}
R.~Dillenschneider, Quantum discord and quantum phase transition in spin
  chains, Phys. Rev. B 78 (2008) 224413.

\bibitem{discord_zero2}
M.~S. Sarandy, Classical correlation and quantum discord in critical systems,
  Phys. Rev. A 80 (2009) 022108.

\bibitem{discord_zero3}
J.~Maziero, H.~C. Guzman, L.~C. C\'eleri, M.~S. Sarandy, R.~M. Serra, Quantum
  and classical thermal correlations in the $\mathit{XY}$ spin-$\frac{1}{2}$
  chain, Phys. Rev. A 82 (2010) 012106.

\bibitem{discord_finite1}
T.~Werlang, C.~Trippe, G.~A.~P. Ribeiro, G.~Rigolin, Quantum correlations in
  spin chains at finite temperatures and quantum phase transitions, Phys. Rev.
  Lett. 105 (2010) 095702.

\bibitem{discord_finite2}
T.~Werlang, G.~A.~P. Ribeiro, G.~Rigolin, Spotlighting quantum critical points
  via quantum correlations at finite temperatures, Phys. Rev. A 83 (2011)
  062334.

\bibitem{rosario_qpt}
A.~Osterloh, L.~Amico, G.~Falci, R.~Fazio, Scaling of entanglement close to a
  quantum phase transition, Nature 416 (2002) 608 EP.

\bibitem{discord_qpt1}
T.~Werlang, G.~Rigolin, Thermal and magnetic quantum discord in heisenberg
  models, Phys. Rev. A 81 (2010) 044101.

\bibitem{discord_qpt2}
Y.-X. Chen, S.-W. Li, Quantum correlations in topological quantum phase
  transitions, Phys. Rev. A 81 (2010) 032120.

\bibitem{discord_oqs}
A.~Shabani, D.~A. Lidar, Vanishing quantum discord is necessary and sufficient
  for completely positive maps, Phys. Rev. Lett. 102 (2009) 100402.

\bibitem{discord_dynamics}
S.~Alipour, A.~Mani, A.~T. Rezakhani, Quantum discord and non-markovianity of
  quantum dynamics, Phys. Rev. A 85 (2012) 052108.

\bibitem{discord_bio}
K.~Br\'adler, M.~M. Wilde, S.~Vinjanampathy, D.~B. Uskov, Identifying the
  quantum correlations in light-harvesting complexes, Phys. Rev. A 82 (2010)
  062310.

\bibitem{coherence_as_resource}
A.~Streltsov, G.~Adesso, M.~B. Plenio, Colloquium: Quantum coherence as a
  resource, Rev. Mod. Phys. 89 (2017) 041003.

\bibitem{old_coherence1}
R.~J. Glauber, Coherent and incoherent states of the radiation field, Phys.
  Rev. 131 (1963) 2766--2788.

\bibitem{old_coherence2}
E.~C.~G. Sudarshan, Equivalence of semiclassical and quantum mechanical
  descriptions of statistical light beams, Phys. Rev. Lett. 10 (1963) 277--279.

\bibitem{baumgratz}
T.~Baumgratz, M.~Cramer, M.~B. Plenio, Quantifying coherence, Phys. Rev. Lett.
  113 (2014) 140401.

\bibitem{coherence_communication}
U.~K. Sharma, I.~Chakrabarty, M.~K. Shukla, Broadcasting quantum coherence via
  cloning, Phys. Rev. A 96 (2017) 052319.

\bibitem{coherence_thermo}
F.~G. S.~L. Brand\~ao, M.~Horodecki, J.~Oppenheim, J.~M. Renes, R.~W. Spekkens,
  Resource theory of quantum states out of thermal equilibrium, Phys. Rev.
  Lett. 111 (2013) 250404.

\bibitem{coherence_bio}
S.~Lloyd, Quantum coherence in biological systems, J. Phys.: Conf. Ser. 302~(1)
  (2011) 012037.

\bibitem{coherence_spin1}
C.~Radhakrishnan, I.~Ermakov, T.~Byrnes, Quantum coherence of planar spin
  models with dzyaloshinsky-moriya interaction, Phys. Rev. A 96 (2017) 012341.

\bibitem{coherence_spin2}
B.~Çakmak, G.~Karpat, F.~Fanchini, Factorization and criticality in the
  anisotropic xy chain via correlations, Entropy 17~(2) (2015) 790–817.

\bibitem{coherence_spin3}
G.~Karpat, B.~\ifmmode~\mbox{\c{C}}\else \c{C}\fi{}akmak, F.~F. Fanchini,
  Quantum coherence and uncertainty in the anisotropic xy chain, Phys. Rev. B
  90 (2014) 104431.

\bibitem{coherence_spin4}
Y.-C. Li, H.-Q. Lin, Quantum coherence and quantum phase transitions, Sci Rep 6
  (2016) 26365 EP.

\bibitem{bethe}
H.~{Bethe}, {Zur Theorie der Metalle}, Z. Phys. 71 (1931) 205--226.

\bibitem{jordan_wigner}
P.~{Jordan}, E.~{Wigner}, {{\"U}ber das Paulische {\"A}quivalenzverbot}, Z.
  Phys. 47 (1928) 631--651.

\bibitem{quspin1}
P.~Weinberg, M.~Bukov, {QuSpin: a Python Package for Dynamics and Exact
  Diagonalisation of Quantum Many Body Systems part I: spin chains}, SciPost
  Phys. 2 (2017) 003.

\bibitem{quspin2}
P.~{Weinberg}, M.~{Bukov}, {QuSpin: a Python Package for Dynamics and Exact
  Diagonalisation of Quantum Many Body Systems. Part II: bosons, fermions and
  higher spins}, ArXiv e-prints\href {http://arxiv.org/abs/1804.06782}
  {\path{arXiv:1804.06782}}.

\bibitem{wells}
H.~J. {Wells}, {Quantum spin chains and random matrix theory}, ArXiv
  e-prints\href {http://arxiv.org/abs/1410.1666} {\path{arXiv:1410.1666}}.

\bibitem{wick}
E.~{Lieb}, T.~{Schultz}, D.~{Mattis}, {Two soluble models of an
  antiferromagnetic chain}, Ann. Physics 16 (1961) 407--466.

\bibitem{concurrence1}
S.~Hill, W.~K. Wootters, Entanglement of a pair of quantum bits, Phys. Rev.
  Lett. 78 (1997) 5022--5025.

\bibitem{concurrence2}
W.~K. Wootters, Entanglement of formation of an arbitrary state of two qubits,
  Phys. Rev. Lett. 80 (1998) 2245--2248.

\bibitem{xmatrix}
C.-Z. Wang, C.-X. Li, L.-Y. Nie, J.-F. Li, Classical correlation and quantum
  discord mediated by cavity in two coupled qubits, J. Phys. B: At. Mol. Opt.
  Phys. 44~(1) (2011) 015503.

\bibitem{coherence_distri}
C.~Radhakrishnan, M.~Parthasarathy, S.~Jambulingam, T.~Byrnes, Distribution of
  quantum coherence in multipartite systems, Phys. Rev. Lett. 116 (2016)
  150504.

\bibitem{jsd}
J.~Lin, Divergence measures based on the shannon entropy, IEEE Trans. Inform.
  Theory 37~(1) (1991) 145--151.

\bibitem{magnetic_ent}
F.~K. Fumani, S.~Nemati, S.~Mahdavifar, A.~H. Darooneh, Magnetic entanglement
  in spin-1/2 xy chains, Phys. A 445 (2016) 256 -- 263.

\bibitem{ent2}
H.~Yano, H.~Nishimori, Ground state entanglement in spin systems, Prog. Theor.
  Phys. Supp. 157 (2005) 164--167.

\bibitem{qc_persist}
A.~Ferraro, L.~Aolita, D.~Cavalcanti, F.~M. Cucchietti, A.~Ac\'{\i}n, Almost
  all quantum states have nonclassical correlations, Phys. Rev. A 81 (2010)
  052318.

\bibitem{sp_coherence}
C.~Radhakrishnan, M.~Parthasarathy, S.~Jambulingam, T.~Byrnes, Quantum
  coherence of the heisenberg spin models with dzyaloshinsky-moriya
  interactions, Sci Rep 7~(1) (2017) 13865.

\bibitem{discrepancy}
M.-L. Hu, H.~Fan, Relative quantum coherence, incompatibility, and quantum
  correlations of states, Phys. Rev. A 95 (2017) 052106.

\bibitem{sachdev}
S.~Sachdev, Quantum Phase Transitions, Cambridge University Press, 2000.

\bibitem{qpt_xx}
T.~{Zhang}, P.-X. {Chen}, W.-T. {Liu}, C.-Z. {Li}, {Classification of
  entanglement and quantum phase transition in XX model}.

\bibitem{qpt_xy}
Y.-C. Li, H.-Q. Lin, Thermal quantum and classical correlations and
  entanglement in the $\mathit{XY}$ spin model with three-spin interaction,
  Phys. Rev. A 83 (2011) 052323.

\end{thebibliography}
\end{document}